\documentstyle[prb,aps]{revtex}
\input psfig.tex
\newcommand{\beq}{\begin{equation}}
\newcommand{\eeq}{\end{equation}}

\newcommand{\z}{\zeta_{sc}}
\begin{document}

 \title{Ostwald ripening in Two Dimensions: 
Correlations and Scaling Beyond Mean Field.}
 \author{Boris Levitan and Eytan Domany}
 \address{Department of Physics of Complex Systems, Weizmann Institute
of Science, Rehovot, Israel}
\date{\today}
\maketitle
\begin{abstract}
We present a systematic quasi-mean field model of the Ostwald ripening process
in two dimensions. Our approach yields a set of dynamic equations for the
temporal evolution of the minority phase droplets' radii. The equations contain
only pairwise interactions between the droplets; these interactions are
evaluated in a mean- field type manner. We proceed to solve numerically the
dynamic equations for systems of tens of thousands of interacting droplets. The
numerical results are compared with the experimental data obtained by
Krichevsky and Stavans for the relatively large volume fraction $\varphi=0.13$.
We found good agreement with experiment even for various correlation functions.
\end{abstract}
 
\section{Introduction}
When a system is quenched into a two-phase coexistence region, its homogeneous
initial state no longer corresponds to thermodynamic equilibrium\cite{RevGunt}. 
If there is a conserved quantity whose density is different in the two
coexisting phases (such as the total volume or total amount of impurities), the
final equilibrium state consists of two macroscopic domains, separated by a
single phase boundary. The manner in which a system evolves from its homogeneous
initial state to the final equilibrium state of two-phase coexistence has been
the subject of numerous theoretical and experimental investigations.

For small initial super-saturation the system's evolution towards two-phase
equilibrium starts by nucleation and growth of droplets (or small crystallites)
of the minority phase.  In the late stage of evolution to the equilibrium state
no new droplets are formed and the amount of material in each of the phases
remains fixed.  Evolution proceeds by means of dissolution of small droplets
and growth of the large ones, giving  rise to reduction of the total (surface)
energy of the system. The exchange of material is driven by diffusion; the
concentration is higher near surfaces with high curvature, and hence the
diffusive flux is directed from the small droplets towards the larger ones. 
This coarsening process\cite{Ostwald}, called {\it Ostwald ripening}, in the
course of which the number of droplets decreases, while the average size of the
remaining ones increases, is the subject of the present paper.

In what follows we describe a formalism that leads to an efficient numerical
algorithm for Ostwald ripening in two dimensions. Our work was motivated by,
and our results are compared with recent experimental work on a two-dimensional
film of liquid and crystalline succinonitrile in coexistence\cite{KS}.
Therefore we will refer to the minority phase as "solid", and to the matrix in
which the droplets are  embedded as "liquid".  

Ostwald ripening belongs to a family of non-equilibrium phenomena that exhibit
evolution to a {\it scaling state}. By a scaling state we mean that there is a
single length scale in the system, which grows with time. For Ostwald ripening
the obvious length scale is the average droplet size $\langle R\rangle$; its
growth with time follows the celebrated\cite{LifshPit} Lifshitz-Slyozov law,
$\langle R\rangle\sim t^{1/3}$. When rescaled by  this changing length, all
statistical characteristics of the system (such as droplet size distribution,
spatial correlations etc) are time-independent. That is, by comparing two
photographs of the system, taken at different (well separated) times but
properly rescaled, one cannot tell from any statistical measurement which
photograph was taken earlier.

A parameter of central importance on which various characteristics of the
system (such as the droplet size distribution) depend is the relative volume
fraction  of the minority phase, $\varphi$. This parameter determines the ratio
of the size of the droplets and the distance between them. Therefore, $\varphi$
controls the extent to which droplets interact with each other; for very small
volume fractions the interaction is weak, and any particular droplet "feels"
the effect of the others only through an effective medium. This observation
served as the basis of the theoretical, mean - field
models\cite{LifshSlyoz,Wagner,Marq} (see ref.\cite{YaoRev} for a review). As
$\varphi$ increases correlations between neighboring droplets become more
important, until mean field based approximations lose their validity. The
effects of correlations have been taken into account analytically to first
order in a small parameter\cite{Marder,ZhGunt}. In three dimensions the small
parameter is $\sqrt{\varphi}$ (see Marder\cite{Marder}) and in two -
$1/\log\varphi$ (see Zheng and Gunton\cite{ZhGunt}).  The obtained expressions
are so complicated even in first order, that it is hard to see how one can
proceed to higher orders. Rather, a combined analytical and numerical approach
seems to be suitable.

The basic phenomenological description of Ostwald ripening is provided in the
framework of a Cahn-Hilliard equation\cite{Cahn} (see ref.\cite{Chakra} for a
review) for the order parameter. Rogers and Desai\cite{RogersDes} performed 
such a
``first principle" computation, reaching the scaling state with about 500
droplets. This was achieved for small values of $\varphi$ (less than 0.1), when
all the droplets were almost circular. Such a number of droplets suffices for
studying the droplet size distribution, but is too small to gain insight into
the spatial and temporal correlations in the scaling state. Moreover, for
larger fractions the final stage of the coarsening process was not achieved in
the calculation and scaling was not observed. 
Masbaum\cite{Masbaum} used this method recently for a simulation, performed on
a parallel computer, starting with about $3000$ droplets. He obtained rather
good agreement for various correlation functions with the experiments of
Krichevsky and Stavans\cite{KS}.

It is clear, however, that this method can hardly be applied for studying much
larger systems, which are needed in order to make statistically meaningful
measurements in the late-stage scaling regime.  Therefore one turns to
simplified descriptions of the system, hoping that the main features of the
full Cahn-Hilliard theory are preserved. For the sake of completeness, we
briefly present below a sequence of approximations that lead from the full
Cahn-Hilliard theory to our approach.

First, one goes to a coarse-grained representation of the order parameter
field, which retains the {\it boundaries} between the crystalline droplets and
the surrounding liquid as one of the dynamic variables of the problem. The
second variable is the diffusing {\it concentration field} (of the impurities
or the liquid itself), $c(\vec r)$. Simplification of the problem is attained
by separation of time scales. The process which occurs on fastest scales is the
equilibration of the concentration near the (moving) boundaries.  Assuming that
this occurs instantaneously allows us to write the Gibbs-Thomson condition for
the concentration at a point on the boundary of a droplet, where the (local)
radius of curvature is $R(t)$; 
\beq
c|_{droplet}(t)=c_{eq}(R(t))=c_{\infty}+{\alpha\over R(t)}.
\label{eq:GT}\eeq
Here $c_{\infty}$ is the equilibrium
concentration in a liquid above a {\it planar} liquid-solid interface.
The problem becomes now one of solving the diffusion equation
\beq
\frac{\partial c}{\partial t}=\nabla^2 c
\label{eq:diff}\eeq
with boundary conditions given by Eq.(\ref{eq:GT}) on {\it moving boundaries}. 

The rate of change of the boundaries is determined by $J_\perp=\hat{n} \cdot
\vec{J}$, the mass flux normal to the droplet's surface, which, in turn, is
proportional to the gradient of the concentration field at the boundary;
\beq
\frac{d R}{d t} = -J_\perp  \qquad \qquad \vec{J} = -\ \nabla c|_R.
\label{eq:J}\eeq

Equations Eq.(\ref{eq:GT}-\ref{eq:J}) constitute a closed dynamical problem. 
The next simplification is a quasi-static approximation to the problem. For
{\it fixed} boundary conditions the diffusing field would reach a steady state
in a characteristic diffusion time $t_D \sim R^2 $, where $R$ is a typical
length scale (distance between neighboring droplets). If this time is much
shorter than the growth time $t_G$, i.e. the time it takes a typical droplet's
radius to change appreciably, the concentration field reaches a steady state
before the radii of the droplets had a chance to change and one can use the
stationary diffusion equation instead of Eq.(\ref{eq:diff}). This is precisely
the case for late stages of the growth process, where $t_G \sim R^3$, so that
indeed $t_D\ll t_G$. The Lifshitz-Slyozov scaling law, $t_G \sim R^3$, can be
established by dimensional analysis of Eqs.(\ref{eq:GT}) and (\ref{eq:J}). 
Hence at late stages the solution $c(\vec{r},t)$ of Eq.(\ref{eq:diff}) can be
approximated by the solution of the stationary diffusion problem 
\beq
\nabla^2 c=0 
\label{eq:Lap} \eeq 
with the previous boundary conditions Eq.(\ref{eq:GT}); but now we look for a
solution of Eq.(\ref{eq:Lap}), from which the rate of change of the boundaries
is obtained using Eq.(\ref{eq:J}), giving rise to new boundaries at the next
time step, and so on. It should be noted that in this approximation  
the total area of the droplets is conserved exactly.

It is possible to reformulate the stationary diffusion problem 
Eqs.(\ref{eq:GT}) - (\ref{eq:Lap}) as an integral equation which automatically
accounts for the boundary conditions. Expanding the shapes of the droplets
using a set of orthogonal polynomials one can reduce the problem to an 
$implicit$ system of ordinary differential equations in terms of the expansion
coefficients\cite{Kawasaki,AkaiwaV,AkaiwaM}.  

Further simplification can be achieved by neglecting the deviation of the
droplet shapes from circular. We expect this approximation to work for low
volume fractions $\varphi$, when the distances between the  droplets are much
larger than the droplets sizes and redistribution of material in a single
droplet is much faster than the exchange between the well separated droplets.
However, experiments\cite{KS} show that even for fractions as large as $\varphi
=0.4$ the droplets are more or less circular. Therefore, we consider Laplace's
equation Eq.(\ref{eq:Lap}) with the boundary conditions Eq.(\ref{eq:GT}) at
perfectly circular domains of radii $R_i$, positioned at points $\vec{r}_i$. 
This is a problem in electrostatics - of calculating the potential in the
presence of conducting cylinders. One approximates the solution of this problem
by 
\beq
c(r)= \sum_i q_i  \log ( |\vec{r} - \vec{r}_i| / R_0 )+
\sum_i {(\vec p_i\cdot(\vec{r}-\vec r_i))\over |\vec{r} - \vec{r}_i|^2},
\label{eq:sol}\eeq
where $R_0$ is an arbitrary length. The "charges" $q_i$ and the "dipoles" $p_i$
should be chosen so that the boundary conditions (\ref{eq:GT}) are satisfied
(approximately) on the surface of each droplet. Let us denote by
$X_{ij}= |\vec{r_i} - \vec{r}_j|$
the distance between the centers of droplets $i$ and $j$, and by $c|_{R_i}$
the correct boundary value of the concentration, as given by Eq.(\ref{eq:GT}).
We now approximate the concentration field at the the boundaries by its 
expansion to first order in the small parameter $R_i/X_{ij}$: 
$$
c_{\infty}+{\alpha\over R_i}= c(\vec r_i + \vec R_i) \approx q_i \log R_i/R_0 + 
{(\vec p_i\cdot\vec R_i)\over R_i^2}~+~~~~~~~~~~~~~~~~~~~~~~$$
\beq
~~~~~~~~~~~~~+~\sum_{j \neq i} q_j \log ( X_{ij}/R_0 )+ 
\sum_{j \neq i} {(\vec p_j\cdot\vec X_{ij})\over |X_{i,j}|^2}+
\sum_{j\ne i}q_j{(\vec X_{i,j}\cdot \vec R_i)\over |X_{i,j}|^2}.
\label{eq:BC}\eeq
$\vec R_i$ denotes the radius-vector of  points on the surface of droplet $i$;
the last term appears as the expansion of $q_j \log |\vec r_i + \vec R_i - \vec
r_j|$. Equation (\ref{eq:BC}) contains two parts; one that depends on the
direction of $\vec R_i$ and one that does not.  This results in two sets of
linear equations for the charges and the dipoles:
\beq
c_{\infty}+{\alpha\over R_i}=q_i \log R_i/R_0 + 
\sum_{j \neq i} q_j \log (X_{ij}/R_0 )+\sum_{j \neq i} 
{(\vec p_j\cdot\vec X_{ij})\over |X_{i,j}|^2},
\label{eq:charge}\eeq 
\beq 
{(\vec p_i\cdot\vec R_i)\over R_i^2}+
\sum_{j\ne i}q_j{(\vec X_{i,j}\cdot\vec R_i)\over |X_{i,j}|^2}=0.
\label{eq:dip}\eeq
As we will see, the last term in Eq.(\ref{eq:charge}) is
$O((R/X)^2)$ - i.e. smaller than the other terms and can be omitted.  Thus the
system of equations 
\beq
c_{\infty}+{\alpha\over R_i}=q_i \log R_i/R_0 + 
\sum_{j \neq i} q_j \log (X_{ij}/R_0 ) 
\label{eq:charges}\eeq
determines the charges.
Since Eq.(\ref{eq:dip}) should be satisfied for any $\vec R_i$, the
dipoles are determined by
\beq
{\vec p_i\over R_i^2}=-\sum_{j\ne i}q_j{\vec X_{i,j}\over |X_{i,j}|^2}.
\label{eq:p}
\eeq
The physical meaning of the Eq.(\ref{eq:p}) is that the dipole associated with
droplet $i$ is simply related to the electric field induced at its center by
all the other charges. Once the charges are obtained from solving Eq.
(\ref{eq:charges}), their substitution in Eq.(\ref{eq:p}) determines the
dipoles.  

The normal component of the flux at the boundary of droplet $i$ is given by
\beq
\vec\nabla_n 
c(\vec r_i + \vec R_i)=\frac{q_i}{R_i}-2{(\vec p_i\cdot\vec R_i)\over R_i^3},
\label{flux}
\eeq
where Eq. (\ref{eq:BC}) and (\ref{eq:p}) were used. The normal flux has two
contributions; an isotropic part, giving rise to a rate of change of the radius,
given by
\beq
\frac{d R_i}{d t} = \frac{q_i}{R_i}
\label{eq:dR}\eeq
and an anisotropic part, due to the dipole term in Eq.(\ref{flux}). The
contribution of the dipole part to the {\it total} flux vanishes; it induces
deposition of material on one side of the droplet and evaporation from the
opposite side.  We approximate the effect of the dipole flux by {\it shifting
the positions} of the circular droplets (see below). We also show (in Appendix
A) that in the low concentration ($\varphi << 1$) limit the effect of the
dipoles is negligible; we first discuss this limit, working with charges only,
and then include the dipoles.

Eq.(\ref{eq:dR}) implies that the rate of change of the area of droplet $i$ is
proportional to its charge $q_i$. Using $q_i$ from Eq.(\ref{eq:dR}) in
Eq.(\ref{eq:charges}) eliminates the concentration field $c(\vec r)$ from the
problem and, after proper rescaling of variables, the following set of
equations for the temporal evolution of the radii results:
\beq
(1+R_i/R_c)=\sum_j L_{i,j} \dot R_j,
\label{eq:dRdt}\eeq
where $R_c=\alpha/c_\infty$ is a capillary length and the matrix $L_{i,j}$ 
defined in 2-$d$ as follows\cite{Kawasaki}:
\beq
L_{i,j} =R_iR_j\cases{\log (R_i/R_0) & if $i=j$\cr
			   \log (X_{i,j}/R_0) & otherwise.\cr}
\label{eq:Lij}\eeq

Clearly, solving the system Eq.(\ref{eq:dRdt})-(\ref{eq:Lij}) is a much simpler
computational task than solving the Laplace problem Eq.(\ref{eq:Lap}) with
moving boundary conditions on the surfaces of the droplets. Note, however, that
the matrix elements $L_{i,j}$ grow with the distance between the droplets. 
Because of this, finite size effects become crucial: all droplets in the bulk
feel the boundary. The simplest way to avoid this problem is to consider the
system with periodic boundary conditions; the interaction of a pair of droplets
contains an infinite sum of logarithms, corresponding to all the $images$ of
these droplets. Yao et al\cite{YaoI} have summed this series using Ewald 
summation techniques. 

The problem is further complicated by the need to invert the $N\times N$ matrix
$L_{i,j}$ at $each$ time step in order to get from Eq(\ref{eq:dRdt}) $explicit$
expressions for $\dot R_i$. Thus, solving the system 
(\ref{eq:dRdt}-\ref{eq:Lij}) takes $N^3$ operations per time step; therefore
each run costs $N^4$ operations. This necessitates huge CPU times (the
simulations of Yao et al\cite{YaoI} took about 1000 CPU hours on an IBM 3090
computer for a system of about $3000$ droplets).  

Beenakker\cite{Been} has solved an analogous problem in 3-$d$ by truncating the
matrix $L_{i,j}$ (defined in 3-$d$ in a way analogous to Eq.(\ref{eq:Lij})) .
He took into account only the interactions between droplets whose separation
does not exceed a threshold, reducing this way the number of droplets with
which a given one interacts to about 20. The physical motivation for such
truncation is {\it screening} \cite{MarderI,MarqRoss}; droplets whose
separation exceeds the screening length do not affect each other. It should be
noted that with such a truncation the total volume of the droplets is not
conserved, and Beenakker had to adjust $R_c$ in Eq.(\ref{eq:Lij})) at each time
step in order to restore volume conservation.

Akaiwa and Meiron\cite{AkaiwaM} used the analogous truncation procedure in
2-$d$. Although the interactions between the charges are expected to be well
screened even in 2-$d$, formal truncation of the matrix seems problematic in
this case.  Since the matrix elements grow with distance between the droplets,
the elements of the $inverse$ matrix as functions of the cutoff should contain
large fast oscillating components (unlike in 3-$d$, where their dependence on
the cutoff is smooth). Nevertheless, their results (in particular, the
correlation functions) compare well with experiment.  Apparently, this success
is due to the fact that these oscillations are effectively averaged out during
the run.  

Our goal was to find an analytic way of approximating $\hat L^{-1}$, in
a manner that reflects screening as the physical basis of the approximation
scheme. Once $L$ has been inverted, we can integrate the equations
\beq
\dot R_i=\sum_j L_{i,j}^{-1}(1+R_j/R_c) 
\label{eq:dotR}\eeq
numerically.
The approximation introduced in this work can be summarized by the expression
\beq
\dot R_i=\sum_{j} L^{-1}_{i,j} = 
\frac{1}{ R_i K_0(R_i/\z)}\sum_{j (\neq i)}
\frac{K_0(X_{i,j}/\z )}{K_0 (R_j /\z)} 
\left( \frac{1}{R_j}-\frac{1}{R_i} \right),  
\label{eq:Linv}\eeq
where $K_0$ is the zeroth order modified Bessel function of the second kind
(also called MacDonald function). Since $K_0(x)\sim\exp{(-x)}$ for large $x$, the
parameter $\z$ has the meaning of a screening length. In the mean field limit
(i.e. for very small area fraction $\varphi$) it is determined by the equation: 
\beq
\z^{-2}=2\pi n\left\langle {1\over K_0(R/\z)}\right\rangle.
 \label{scr}\eeq
However, for larger fractions, where the effects of correlations become
important, $\z$ is determined by a more complicated condition, as discussed in
Appendix B. In practice, in our simulations for $\varphi=0.13$ we used
$\z=2.73\bar R$, as determined by Table I (see Appendix B for details).

In spite of various approximations that were made to derive Eqs.(\ref{eq:Linv})
- (\ref{scr}), we believe that they do contain the most important aspects of
the dynamics present in Eq.(\ref{eq:dRdt}). One can see explicitly the effects
of screening, with the screening length appearing naturally in the derivation
presented below. Total area is conserved explicitly. In the limit of very small
minority phase area fractions, where $\langle R\rangle/\z\rightarrow 0$, our
description is consistent with Marqusee's mean-field theory\cite{Marq}, which
leads to the following dynamical equation for the droplet's radius:
\beq
R{dR\over dt} = k(R/\zeta)\left({1\over R}-{1\over R_c}\right),
\label{eq:Marqusee}\eeq
where $k(x)={xK_1(x)~/~K_0(x)}$, while $\zeta$ is defined by the equation:
$\zeta^{-2}=2\pi n\left\langle k(R/\zeta)\right\rangle$.
Since $xK_1(x)\rightarrow 1$ for $x\rightarrow 0$, we find $k(x)\rightarrow
1/K_0(x)$ and our definition of the screening length coincides with his.  

We used the formalism presented above to integrate the evolution of large
assemblies of droplets and found that at relatively large values of the area
fraction $\varphi$ the droplets' motion, induced by the so far neglected
dipoles becomes important and must be taken into account.  

We evaluate now the contribution of the dipoles to the shift of the droplets'
centers of mass, $\delta\vec r_i$, defined by 
\beq
M\delta\vec r = \int\vec R\delta m.
\label{deltaX}\eeq
Here the integration is over the boundary of a droplet; $\vec R$ is a radius
vector on its boundary, $M=\rho\pi R^2$ is the total mass of the droplet
($\rho$ is the density; it will drop out of the final result) and $\delta m$ is
the additional mass adsorbed (or lost) locally, due to the shift of the
boundary during the interval $dt$. The anisotropic component of the local
velocity of the droplet's boundary is given by
$$v=-J_n=-{2p\cos\theta\over R^2},$$
where $\theta$ is the polar angle between $\vec R$ and $\vec p$. The mass added 
at $\vec R$ is
$$\delta m =\rho(vdt)(R~d\theta)=-(2p/R)\rho dt\cos\theta~d\theta,$$
so that (see also Appendix A of \cite{Marder})
$${d\vec r\over dt}=-{2p\over \pi R^3}
\int\vec R \cos\theta~d\theta=-{2\vec p\over\pi R^2}\int\cos^2\theta~ 
d\theta =-{2\vec p\over R^2}. $$
Finally,
\beq
{d\vec  r_i\over dt} =-{2\vec p_i\over R_i^2}=
2\sum_{j\ne i}{q_j\over |X_{i,j}|}{\vec X_{i,j}\over |X_{i,j}|}.
\label{shift}\eeq
Note, that both sides of Eq.(\ref{shift}) have the same dimensionality, since
the charges $q_i$ are measured in area per time (see Eq.(\ref{eq:dR})). 

The procedure for solving Eq.(\ref{eq:dRdt})-(\ref{eq:Lij}) for the dynamics of
the droplets' radii $R_i$, together with Eq.(\ref{shift}) for their positions
$\vec r_{i}$ is as follows.  For given $\vec r_{i}$ and $R_i$ we invert $\hat
L$, obtain $dR_i/dt$ (or $q_i$ as given by Eq(\ref{eq:dR})) and integrate one
time step.  Next, substituting $q_i$ into Eq.(\ref{shift}) we obtain new values
of $\vec r_i$ and the procedure is repeated.

The sum on the r.h.s. of Eq.(\ref{shift}) appears to be problematic: assuming
that the droplets' charges are uncorrelated (while the total charge is zero)
one can see that the mean square of this sum  diverges logarithmically with the
size of the system. However, as we show in Appendix A, the correlations between
the charges, provided by screening, ensure the convergence of this sum,
so that it can be evaluated using  a reasonable fraction of its terms.
Actually, a {\it heuristic} formula, simpler than that of  Eq.(\ref{shift}),
can be used to evaluate  the droplets' shifts in practical simulations. As
explained in Appendix A, Eq.(\ref{shift}) can be replaced by 
\beq
{d\vec r_i\over dt} =2\sum_{j\ne i}
\frac{K_{j,i}}{k_ik_j} 
\left( \frac{1}{R_i}-\frac{1}{R_j} \right)
{\vec X_{i,j}\over |X_{i,j}|^2},
\label{Move}
\eeq
where $K_{i,j}=K_0(X_{i,j}/\zeta_{sc})$ and $k_i=K_0(R_i/\zeta_{sc})$. 
Although the approximation leading to this formula is not clearly established,
our simulations show that it works as well as the more rigorous
Eq.(\ref{shift}). At the same time it is much more economic because only
droplets that lie within the screening length (from the droplet whose shift is
evaluated) contribute to the sum.

In the sequence of approaches to the Ostwald ripening problem in two
dimensions, ranging from the Cahn-Hilliard equation to Marqusee's mean field
theory, our model, which includes only the pairwise interaction between the
droplets, can be viewed as the {\it minimal extension} of the mean field
approach. This concerns our Eq.(\ref{Move}) (in comparison with the
Eq.(\ref{shift})) as well. We believe that our approach constitutes the
simplest step that can be taken beyond simple mean field.  

Our program is as follows. In the next section we present a few universal
properties of the matrix $L$. This is followed by Section III, where we derive
the main result, Eq.(\ref{eq:Linv}), using a mean-field type approximation to
$L^{-1}$.  Numerical solutions of the resulting dynamics, based on
Eq.(\ref{eq:Linv}), and on both (\ref{shift}) or (\ref{Move}), are presented in
Section IV and compared with experiments by Stavans and Krichevsky. Section V
presents a short summary of our approach and results.

\section{Some Sum Rules}
We present now some important properties of the matrix $L^{-1}$, that lead
to useful ``sum rules". The first claim is that
\beq
\sum_i L_{i,j}^{-1}R_i=\sum_j L_{i,j}^{-1}R_j\rightarrow 0
\label{eq:sum0}\eeq
the sum approaching zero as the system size increases. This relationship is
analogous to one presented for the 3-$d$ case by Beenakker and
Ross\cite{BeenRossI}. We present here arguments (that seem to us simpler than
those of ref.\cite{BeenRossI}) for the validity of Eq.(\ref{eq:sum0}) in 2-$d$.
First consider the quantity
\beq
a_j=\sum_i L_{i,j}/R_j
=\sum_i R_i\log (X_{i,j}/R_0).
\label{eq:aj}\eeq
The sum is clearly dominated by regions that are far from $\vec{r}_j$; in the
absence of long-range correlations the composition of such regions does not
depend on the identity of droplet $j$. On the other hand, the contribution of
those droplets $i$ that lie near droplet $j$ {\it will}, in general, depend on
$R_j$. By ``near" we mean the region for which correlations {\it are}
important. The size of this region is, however, small (on the order of the
screening area - see below), the values of $X_{ij}$ in this region are small,
and the contribution from it is negligible compared to those from the far-away
parts of the plane. That is, for increasing system size $a_j \rightarrow
\infty$, whereas the contribution from the correlation region remains finite. 
Hence we can write 
\beq
a_j=\sum_i L_{i,j}/R_j \simeq a \rightarrow \infty.
\label{eq:a}\eeq
Next, consider the following sum:
\beq
1=\sum_k\delta_{i k}= \sum_k\sum_j L_{i,j}^{-1}L_{j,k}=\sum_j L_{i,j}^{-1}
R_j\sum_kL_{j,k}/R_j=\sum_j L_{i,j}^{-1}R_ja_j
\label{eq:long}\eeq
using now Eq.(\ref{eq:a}) we get
\beq 
a\sum_j L_{i,j}^{-1}R_j \simeq 1
\label{eq:1}\eeq
so that
\beq
\sum_j L_{i,j}^{-1}R_j \simeq a^{-1} \rightarrow 0
\label{eq:00}\eeq
and hence statement Eq.(\ref{eq:sum0}) holds in the limit of large system size.
An immediate consequence of Eq.(\ref{eq:00}) is obtained by using it in 
Eq.(\ref{eq:dotR}), yielding 
\beq
\dot R_i=\sum_j L_{i,j}^{-1}.
\label{eq:Rdot}\eeq
Thus we see that the length $R_c$
drops out from the description of the dynamics of the system.

Multiplying by $R_i$ and summing over $i$, and using again Eq.(\ref{eq:00}),
we establish area conservation
\beq
\sum_i R_i\dot R_i=\sum_j \left( \sum_i L_{i,j}^{-1}R_i\right)=0
\label{eq:area}\eeq
as a consequence of our sum rule. Another important consequence is the
following observation: $L_{ij}^{-1}$, the elements of the inverse matrix,
are {\it independent of the parameter} $R_0$. To see this, take the derivative
of $(L^{-1} L)_{i,k}$ with respect to $\log R_0$: clearly $(L^{-1})^\prime L
+ L^{-1} L^\prime = 0$ and hence
\beq
(L_{i,m}^{-1})^\prime = -\sum_{jk} L^{-1}_{ij} L_{jk}^\prime L^{-1}_{km}.
\label{eq:prime}\eeq
But from Eq.(\ref{eq:Lij}) we immediately get that $(L_{j,k})^\prime=-R_jR_k$,
which when used in Eq.(\ref{eq:prime}) gives
\beq
(L_{i,m}^{-1})^\prime=-\sum_j\sum_k (L_{k,m}^{-1})^\prime R_kL_{i,j}^{-1}R_j
\label{eq:pr}\eeq
and when Eq.(\ref{eq:00}) is used here, we get 
\beq
(L_{i,m}^{-1})^\prime= 0
\label{eq:R00}\eeq
and hence $L^{-1}_{ij}$ does not depend on $R_0$; and in view of
Eq.(\ref{eq:Rdot}) the system's dynamics is, therefore, also independent of
$R_0$.

\section{Mean field  approximation to $L^{-1}$}
Let us  decompose the matrix $\hat L$ of Eq.(\ref{eq:Lij}) to its diagonal and 
off-diagonal parts, 
\beq
{\hat L} =  {\hat L}_0 - {\hat L}_1,
\label{eq:diag0}\eeq
where $(\hat L_0)_{i,j}=\delta_{i,j}R_i^2\log (R_i/R_0)$ and $(\hat
L_1)_{i,j}=-R_iR_j\log(X_{i,j}/R_0)$. Note, that only the off-diagonal part 
contains interactions between different droplets. One could consider an 
expansion of ${\hat L}^{-1}$ in powers of the interaction:
\beq
\hat L^{-1}=(1-\hat L_0^{-1}\hat L_1)^{-1}\hat L_0^{-1}=\sum_{n=0}^{\infty} 
(\hat L_0^{-1}\hat L_1)^n\hat L_0^{-1}
\label{eq:series}\eeq 
with $\hat L_0^{-1}$ given by
\beq
(\hat L_0^{-1})_{i,j}=\delta_{i,j}{1\over R_i^2\log (R_i/R_0)}.
\label{eq:G0}
\eeq 
Clearly, for matrix elements that correspond to well separated droplets ( i.e. 
when $X_{ij}\gg R_0$), terms of order $n+1$ will be larger than those of order
$n$: that is, the series diverges formally and in order to obtain meaningful
results one should perform some kind of partial summation to all orders. To do
this we introduce the $T-matrix$, defined by
\beq
\hat L^{-1}=\hat L_0^{-1}
+\hat L_0^{-1} \hat T \hat L_0^{-1}.
\label{eq:LinvT}\eeq 
To ensure that 
$\sum_j L_{i,j} L_{j,k}^{-1}=\delta_{i,k}$,
$~~\hat T$ has to satisfy the equation
\beq 
\hat T=\hat L_1+\hat L_1\hat L_0^{-1}\hat T.
\label{eq:T}\eeq 
For the sake of convenience we introduce a new matrix $\hat \phi$, defined by
\beq 
T_{i,j}=R_iR_j\phi_{i,j}.
\label{eq:phi}\eeq
Once the matrix $\phi_{i,j}$ has been found, it is straightforward to write
down $\hat T$ and, using Eq.(\ref{eq:LinvT}), the inverse matrix $\hat L^{-1}$. 
Rewriting Eq.(\ref{eq:T}) in terms of $\hat \phi$ we get
\beq 
\sum_{j (\neq i)}{1\over \log (R_j/R_0)} \log (X_{i,j}/R_0) \phi_{j,k}+
\phi_{i,k}= -(1-\delta_{ik} )\log (X_{i,k}/R_0).
\label{eq:expl}\eeq
For the diagonal elements this takes the form
\beq
\phi_{k,k}= 
-\sum_{j \neq k}{1\over \log R_j/R_0} \log (X_{k,j}/R_0) \phi_{j,k},
\label{eq:exd}\eeq
whereas for the off-diagonal elements we obtain from Eq.(\ref{eq:expl})
\beq
\sum_{j \neq i,k}{1\over \log R_j/R_0} \log (X_{i,j}/R_0) \phi_{j,k}+
\phi_{i,k}= -\gamma_k\log X_{i,k}/R_0
\label{eq:ex}\eeq
where
\beq
\gamma_k=1+{\phi_{k,k}\over\log(R_k/R_0)}.
\label{eq:gam}\eeq
It should be noted that equations (\ref{eq:exd}-\ref{eq:gam}) are exact. 
In principle for any configuration of droplets, the set of  equations 
(\ref{eq:expl}) are to be solved for the matrix elements $\phi_{i,k}$. 
We approximate the solution of this problem in a {\it mean-field} spirit.
The central premise of this mean field approach is that the off-diagonal
matrix elements $\phi_{i,j}$ depend only on the distance $X_{i,j}$, i.e.
\beq
\phi_{i,j} = 
\phi( |\vec{r_i} - \vec{r_j}|).
\label{eq:phir}\eeq
That is, we are interested only in the $pairwise$ interaction between the 
droplets (neglecting any possible dependence of $\phi_{i,j}$ on other 
droplets). The analogous approximation for the diagonal elements is that they 
are all equal, i.e. $\phi^{mf}_{k,k}$ is independent of $k$;
\beq
\phi_{k,k} \approx \phi^{mf}_{k,k} = \phi_0.
\label{eq:phikkmf}\eeq 
To obtain  manageable mean-field equations the following simplifying 
assumptions are made:
\begin{enumerate}
\item
In sums such as (\ref{eq:exd}-\ref{eq:ex}) replace  $1/ \log (R_j/R_0)$
by its average value\cite{NoteO}.
\item
Approximate these sums by integrals;
\item
Set in Eq. (\ref{eq:ex})
$\gamma_k=\gamma$ for all $k$ (i.e. impose independence of $k$).
\end{enumerate}
We will discuss and justify these steps below, but before doing that, we
investigate the resulting approximation to Eq.(\ref{eq:ex}), given by
\beq
-{1\over 2\pi}\zeta_{0}^{-2} \int\log (X_{i,j}/R_0)\phi(X_{j,k})d^2r_j+
\phi(X_{i,k})=-\gamma \log X_{i,k}/R_0,
\label{eq:naiv}\eeq
where
\beq  
\zeta_{0}^{-2}=2\pi n\left\langle {1\over\log (R_0/R)}\right\rangle
\approx 2\pi n {1\over\log (R_0/\left\langle R \right\rangle)}
\label{eq:z}\eeq
and the angular brackets denote averaging over the distribution of droplet
sizes (that is, the average value obtained in the particular droplet
configuration in which ${\hat L}^{-1}$ is evaluated). 

The integral equation (\ref{eq:naiv}) can be solved easily by operating with 
$\bigtriangledown_i^2$ on the left and the right hand sides. Using the
identity $\bigtriangledown^2\log r=2\pi\delta({\vec r})$ we obtain the
following differential equation for $\phi(x)$:
\beq  
-\zeta_{0}^{-2}\phi+\bigtriangledown^2\phi=-2\pi \gamma \delta({\vec r}).
\label{eq:Bes}\eeq 
Its solution is the MacDonald function, 
\begin{eqnarray} 
\phi_{i,k} &=&\phi(X_{i,k})= \gamma K_0(X_{i,k}/\zeta_{0}) ~~~~~~i \neq k   
\label{eq:phiik} \\
\nonumber \\
K_0(r/\zeta_{0})
 &\approx & \cases{ -\log ( r /2 \zeta_0 )-C  & \qquad
 $r\ll\zeta_0$  \cr \cr
\sqrt{\frac{\pi \zeta_0}{2 r}} e^{-r/\zeta_0} & \qquad $r \gg \zeta_0$ \cr}
\label{eq:zero}
\end{eqnarray} 
where $C \approx 0.5772$ is Euler's constant.

Let us discuss briefly this result. First of all note that self consistency of
the approximation imposes positivity of $\zeta_0^2$, i.e. $R_0 \gg \langle R
\rangle $. This is indeed the case, as will be discussed below. 
Next note that the
divergence of $\phi(r)$ at short distances is an artifact of the
approximations we made by replacing the exact discrete equation (\ref{eq:ex})
by the continuous Eq.(\ref{eq:naiv}). This divergence has no physical
consequences, since the diagonal elements of the {\it discrete} matrix $\hat
\phi$ are not given by taking $r \rightarrow 0$ in the solution of
Eq.(\ref{eq:Bes}); rather, they are determined by Eq.(\ref{eq:exd}).

One should note also that if $\gamma_k$ were to depend on $k$
the result (\ref{eq:phiik}) would have become 
$\phi_{i,k} = \gamma_k K_0(X_{i,k}/\zeta_{0})$, 
implying that $\phi_{i,k} \neq \phi_{k,i}$,
whereas the matrices ${\hat T}$ and $\hat \phi$ must be symmetric (see Eqs.
(\ref{eq:LinvT}) and (\ref{eq:phi}) and remember that $\hat L$ is symmetric).
 
The diagonal elements are expressed in terms of the off-diagonal ones
in Eq.(\ref{eq:exd}); using there the approximations listed above and 
the mean-field expression (\ref{eq:phiik}) for the off-diagonal elements,
Eq.(\ref{eq:exd}) becomes
\beq
\phi_{k,k} \approx
{1\over 2\pi}\zeta_{0}^{-2}
\int\log (X_{k,j}/R_0)K_0(X_{j,k}/\zeta_0)d^2r_j=\phi_0
\label{eq:phi0}
\eeq
as anticipated in Eq.(\ref{eq:phikkmf}). Using this in Eq.(\ref{eq:gam}) yields
\[
\gamma_k = 1 + \frac{\phi_0}{\log(R_k/R_0)}
\]
and in order to get a consistent mean field approach, with $\gamma_k$
independent of $k$ (as required by the assumption No. (3) from our list), we
must have $\phi_0=0$.  To set $\phi_0=0$ we will now use our freedom to adjust
the parameter $R_0$.  As discussed in the Section II, we are free to vary this
parameter, since it does not affect the dynamics\cite{NoteI}. Thus we arrive at
the following condition on $R_0$:
\beq
\phi_0={1\over 2\pi}\zeta_{0}^{-2}
\int\log (X_{k,j}/R_0)K_0(X_{j,k}/\zeta_0)d^2r_j= 0.
\label{eq:diag00}\eeq
After some simple but lengthy transformations, given in Appendix C, Eq. 
(\ref{eq:diag00}) becomes the following approximate expression for $\log R_0$:
\beq
\left\langle {1\over\log (R_0/R)}\right\rangle 
\approx \left\langle {1\over K_0( R /\zeta_0)}\right\rangle.
\label{eq:con}\eeq
Substituting Eq.(\ref{eq:con}) in Eq.(\ref{eq:z}) yields a closed equation for 
$\zeta_0$, with no dependence on $R_0$:
\beq
\zeta_{0}^{-2} = 2\pi n\left\langle {1\over\log (R_0/R)}\right\rangle 
\approx 2\pi n\left\langle{1\over K_0( R /\zeta_0)}\right\rangle.
\label{eq:z0}
\eeq
Note that once we find $\zeta_0$, we can use this equation also to determine
$R_0$. This will not be necessary since $R_0$ drops out of the dynamic
equations, as shown below. Equation (\ref{eq:z0}) is almost identical to
Marqusee's expression for the screening length. In principle we can now proceed
as planned; for any droplet configuration evaluate $\zeta_0$, calculate the
mean field approximation to the matrix $\hat\phi$, 
\beq 
\phi_{i,j} \approx K_0(X_{i,j}/\zeta_0) \qquad \qquad
\phi_{k,k} \approx 0 
\label{eq:phimf}\eeq
substitute in Eqs.(\ref{eq:phi}) and (\ref{eq:LinvT}) to get ${\hat L}^{-1}$
and use it in the dynamic equation (\ref{eq:Rdot}). There is a problem with
doing this, though. An important property of the {\it exact} ${\hat L}^{-1}$ is
that it satisfies the sum rule (\ref{eq:00}) and hence the total area of all
droplets is conserved by the dynamics (see Eq.(\ref{eq:area})). Since
Eq.(\ref{eq:phimf}) is an approximation, we have to check the extent to which 
the sum rules are satisfied by our approximate $\hat\phi$.
Substituting (\ref{eq:phimf}) in (\ref{eq:LinvT}) yields 
\beq
{\hat L}^{-1}_{i,i} = \frac{1}{R_i^2 \log (R_i/R_0)}, \qquad \qquad
{\hat L}^{-1}_{i,j} = 
\frac{K_0(X_{i,j}/\zeta_0)}{R_i \log (R_i/R_0)~ R_j \log (R_j/R_0)}.
\label{eq:Linvmf}\eeq
Note that ${\hat L}^{-1}_{i,j}$ are small at distances $X_{i,j}\gg\zeta_{0}$;
hence the parameter $\zeta_{0}$ should be interpreted as a {\it screening
length}. At short distances the solution of Eq.(\ref{eq:Bes}) behaves like the
lowest order approximation (i.e. stopping at $n=1$ in the expansion
(\ref{eq:series})). Using these expressions in the sum rule (\ref{eq:00}) gives 
\beq
\sum_j L_{i,j}^{-1}R_j \approx \frac{1}{R_i \log (R_i/R_0)} \left[ 1 +
\sum_{j \neq i} \frac{1}{\log (R_j/R_0)} K_0(X_{i,j}/\zeta_0) \right] 
\label{eq:summf1}\eeq
When we  approximate the sum by an integral, in the spirit of our mean-field 
approach, it becomes
\[
\sum_{j \neq i} \frac{1}{\log (R_j/R_0)} K_0(X_{i,j}/\zeta_0) 
\rightarrow -\frac{1}{2 \pi} \zeta_0^2 \int K_0 (r/\zeta_0) d^2r = -1
\]
and the right hand side of (\ref{eq:summf1}) vanishes, so that in the
mean-field limit (i.e. for vanishing area fraction; see Appendix D)
when our approximations become exact, the sum rule is indeed satisfied. If, 
however, we use Eq.(\ref{eq:Linvmf}) for practical calculations at a 
non vanishing area 
fraction, the sum rule (and hence area conservation) will {\it not} be 
satisfied rigorously.

This problem can be overcome rather simply in the following way. Adopt the
mean-field approximation (\ref{eq:Linvmf}) for the {\it off-diagonal} elements
of ${\hat L}^{-1}$, and use the sum rule (\ref{eq:00}) to determine its
diagonal elements:
\[
{\hat L}^{-1}_{i,i} = -\frac{1}{R_i} \sum_{j \neq i} L_{i,j}^{-1}R_j
\]
The result for both diagonal and off-diagonal elements 
of ${\hat L}^{-1}$ can be summarized as 
\beq
\hat L^{-1} =
\cases{ ~{-1\over R_i^2\log (R_i/R_0)}
\sum_{j (\neq i)} { K_0(X_{i,j}/\zeta_0)\over\log (R_j/R_0)}
& ~~~~~if i=j\cr
{K_0(X_{i,j}/\zeta_0)\over R_iR_j\log (R_i/R_0)~ \log
(R_j/R_0)}& ~~~~~~~otherwise.\cr}
\label{eq:main}
\eeq
With this approximation for ${\hat L}^{-1}$ the sum-rule is of course exactly
satisfied in every step and area is conserved. In the mean-field limit the sum
in Eq.(\ref{eq:main}) yields $-1$, and the expression for ${\hat L}^{-1}_{i,i}$
obviously reduces to the naive one, given by Eq.(\ref{eq:Linvmf}).

Using Eq.(\ref{eq:main}) in Eq.(\ref{eq:Rdot}) we obtain now a rather elegant
expression for $\dot R_i$, the rates of change of the droplets' radii:
\beq
{\dot R}_i=\sum_{j} L^{-1}_{i,j}=
{1\over R_i\log (R_i/R_0)}\sum_{j (\neq i)}
{K_0(X_{i,j}/\zeta_0)\over\log (R_j/R_0)} \left({1\over R_j}-{1\over
R_i}\right).
\label{eq:vell}\eeq
For a given droplet configuration we have now an explicit expression for the
dynamics of the droplets' radii (note that the values of the parameters $R_0$
and $\zeta_0$ are also determined by the configuration). As a final "cosmetic"
adjustment, we use Eq.(\ref{eq:con}) to replace logarithms by $K_0$ (thereby
eliminating $R_0$ from the dynamics), yielding the equation we used in our
numerical study:
\beq
{\dot R}_i=
{1\over R_i K_0(R_i/\zeta_0)}\sum_{j (\neq i)}
{K_0(X_{i,j}/\zeta_0)\over K_0 (R_j/\zeta_0)} \left({1\over R_j}-{1\over
R_i}\right).
\label{eq:vel1} 
\eeq
To complete the treatment we now justify our naive approach, discuss the regime
in which our approximation is expected to hold, and check whether various
assumptions that were made are self-consistent. First of all we assumed that
$R_0$ can be chosen so that  $\zeta_{0}^{2}>0$. Now we can see from Eq.(\ref{eq:con}) that the "optimal"
$R_0\approx\zeta_{0}$, and since in the mean-field limit $\zeta_0\gg R_i$ this
means that $R_0\gg R_i$ for all $i$, so that by Eq.(\ref{eq:z})
$\zeta_{0}^{2}$ is indeed positive and our approach is self-consistent.  

The central point of our approach was replacement of the sum Eq.(\ref{eq:ex})
by an integral: 
\beq 
\sum_{j\neq i,k}{1\over \log R_j/R_0} \log (X_{i,j}/R_0)
\phi_{j,k} \rightarrow n\left\langle {1\over \log R_j/R_0}\right\rangle
\int\log(X_{i,j}/R_0)\phi(X_{j,k})d^2r_j 
\eeq
It is intuitively clear that such a substitution is valid as long as
$N_{\zeta}$, the number of droplets in the screening zone is large. Now, after
having obtained the solution $\phi$, we can indeed show that the condition
$N_{\zeta} \gg 1$ holds when $\log (\varphi^{-1})\gg 2\pi$ (see Appendix D).
Therefore the conditions for validity of the mean-field result lead to a
self-consistent theory only in the limit of very low area fraction,
$\varphi\rightarrow 0$; even for $\varphi\sim 0.001$, there are only a few
droplets in the screening zone.  

For the experimental value $\varphi=0.13$, $\zeta_0$  is about the mean droplet
radius, which makes our approximation completely invalid. Such a small value of
$\zeta_0$ appears because above we have neglected $all$ correlations between
the positions of the droplets. In particular, it turns out that each of the
droplets is surrounded by a {\it depletion zone}, from which all possible
neighbors are excluded (at least, the distance between the centers of any two
droplets must exceed the sum of their radii). In practice (as seen from our
numerical solutions - see below) the mean diameter of the depletion zones $d$
is approximately $2.2\langle R\rangle$. The corrections provided by including
the effects of the depletion zones are discussed in Appendix B.  As is shown
there, taking the depletion zones into account does not change the previously
obtained expression for (\ref{eq:Linv}), but the mean-field $\zeta_0$ is
replaced by a larger screening length ($\z=2.73\langle R \rangle$ for
$\varphi=0.13$), which becomes comparable to the nearest neighbors' distance. 
Although this improvement is not sufficient to justify our mean-field approach,
the numerical results (obtained with the effect of depletion zones taken into
account) presented in the next section are in a rather good agreement with the
experiments that were performed at $\varphi=0.13$.  

\section{The algorithm and  results of the simulations.}
In this section we describe our numerical procedure and the results of our
simulations.  Since we would like to reach the scaling state with a
sufficiently large number of droplets (to reduce the effect of fluctuations),
we must start from initial states with very many droplets and allowing this
large system to evolve for long times. In order to do this with reasonable
computational resources one has to resort to approximations that accelerate the
numerical procedure. Rewrite equations (\ref{eq:vel1}) in the following form:
\beq
{dS_{\alpha}\over dt}=V_\alpha \equiv \sum_\beta v_{\alpha,\beta}
\label{eq:new0}
\eeq
\beq
v_{\alpha,\beta}=
{K_{\alpha,\beta}\over k_{\alpha}~k_{\beta}} \left({1\over
R_{\beta}}-{1\over R_{\alpha}}\right), 
\label{eq:new1}
\eeq
where we used the abbreviated notation 
\[
K_{\alpha,\beta}\equiv
K_0(X_{\alpha,\beta}/\z) \qquad k_\alpha \equiv 
K_0(R_{\alpha}/\z) \qquad S_\alpha \equiv R_\alpha^2/2.
\]
The screening length that appears in the argument of the MacDonald functions
is $\z = 2.73 \bar R$, as follows from the analysis of the depletion zones 
presented in the Appendix B (see Table I at $\varphi=0.13$).

Say we wish to integrate (\ref{eq:new0})-(\ref{eq:new1}) naively. For each time
step we have to evaluate all $N(t)$ velocities $V_\alpha$, with each velocity
given by a sum of $N$ terms, i.e. $N_{op}\approx N^2$ operations per time step.
If we have initially $N$ droplets, and wish to reach the late stages with a few
$\%$ surviving, we would have to perform $N_{steps}\sim N$ time steps of
integration, as was done by Yao et al\cite{YaoI}. This is so because if the
time step is greater than the life time of some droplet, its area will become
$negative$ at the end of the time step, which leads to an increase of the area
fraction of the surviving droplets. Therefore, apparently in order to guarantee
exact area conservation one must eliminate each shrinking droplet separately,
restricting the time step by the next vanishing. At the same time, the detailed
evolution of the smallest droplet is absolutely unimportant for us. Rather we
would like to choose $\tau$ to be not smaller than it is needed to ensure that
the relative change of $\langle R \rangle $, the $mean$ radius of the droplets,
is small during one time step. For this reason, Akaiwa and Meiron\cite{AkaiwaM}
simply removed the droplets with $R<\epsilon\langle R\rangle$ with
$\epsilon=0.1$ $before$ performing the time step. Let $f_s$ be the fraction of
the droplets that are eliminated in each time step (it is kept constant, to a
good approximation, by fixing $\epsilon$). Then the number of steps needed for a
run is reduced to $N_{steps}\approx (1/f_s)\log N$.  However, their procedure
still did not guarantee precise area fraction conservation: when the number of
droplets got reduced by a factor of 5 (from 100 000 to 20 000), the change of
$\varphi$ was about 2\%\cite{YaoI}.  

We require our algorithm to conserve area fraction $exactly$ and, nevertheless,
to reduce both $N_{steps}$ and $N_{op}$. We achieved $N_{op}\sim N$ and
$N_{steps}\sim (1/f_s)\log N$; that is, our scheme requires $O(N\log N)$
operations for the entire evolution.

To reduce the $N$-dependence of $N_{op}$ we note that a droplet $\beta$, whose
separation from $\alpha$ exceeds considerably the screening length, will have
only a small contribution to $V_\alpha$. As will be shown below, only droplets
from the {\it first layer} near $\alpha$ have a significant contribution and
therefore only the neighbors of $\alpha$ are included in the sum $V_\alpha =
\sum_\beta v_{\alpha,\beta}$. Therefore we have $N_{op} \sim N$ 
(replacing $N^2$).  

To reduce $N_{steps}$ to $N_{steps}\approx (1/f_s)\log N$, {\it conserving the
area fraction}, we should treat accurately the vanishing of the small droplets,
instead of simply removing them before performing the time step, as is done in
ref.\cite{YaoI}. Putting this into practice requires care, however.  Let us
consider the system of droplets at two consecutive times $t_0$ and
$t_1=t_0+\tau$. We call all droplets that survive to  $t_1$ "large" and denote
them by indices $i$ and $j$, whereas droplets that vanished before $t_1$  are
called "small" and marked by indices $n,m$. With this notation Eqs.
(\ref{eq:new0})-(\ref{eq:new1}) become
\beq
{dS_i\over dt}=\sum_j v_{i,j}+\sum_m v_{i,m},
\label{eq:new1a}\eeq
\beq
{dS_m\over dt}=\sum_n v_{m,n}+\sum_jv_{m,j}.
\label{eq:new1b}\eeq
We choose $\tau$ to be small enough so that the density of small droplets is
small; hence the typical distance between two small droplets exceeds
considerably the screening length. Therefore to an excellent approximation two
small droplets do not interact (i.e. $K_{m,n} \approx 0$) and the first term in
(\ref{eq:new1b}) can be neglected. Thus Eq.(\ref{eq:new1b}) becomes 
\beq
{dS_m\over dt}=\sum_j v_{m,j} = V_m.
\label{new1bb}\eeq
Then a small droplet will live for time 
\beq
t_m = -S_m/V_m.
\label{tm}\eeq
Turning now to integrate Eq.(\ref{eq:new1a}) we note that the term $v_{i,m}$,
which represents the contribution of the fast droplet $m$ to the growth of the
slow droplet $i$ is actually present only during the interval $t_m$ and not
the entire $\tau$. Hence we must use
\[
S_i(t+\tau) = S_i(t) + \left(\sum_j v_{i,j} \right) \tau + \sum_m v_{i,m} t_m
\]
which can be rewritten, using Eq.(\ref{tm}), as
\beq
S_i(t+\tau) = S_i(t) + \left(\sum_j v_{i,j} + 
\frac{1}{\tau}\sum_m S_m \frac{v_{i,m} }{V_m}\right)\tau.
\label{eq:new1aa}\eeq
This is, in fact, the discrete-time form of the differential equation
\beq
{dS_i\over dt}=\sum_j v_{i,j} + {1\over\tau}\sum_m S_m{v_{m,i}\over V_m}.
\label{Final}\eeq
This equation has a very simple interpretation: the $i$-th large droplet
obtains from each of the small droplets a part of the latter's area,
proportional to the strength of the interaction between the droplets. 
Eq.(\ref{eq:new1aa}) is the main working formula of our algorithm. The two
terms in the parentheses correspond to (i) redistribution of the material
between the large droplets and (ii) absorption of the material that leaves the 
small droplets onto the large ones. Using Eq.(\ref{eq:new1aa}) eliminates the
fast scale of the dynamics; the time scale $\tau$ is to be chosen so that the
assumptions made above are indeed satisfied. Our scheme is put to practice by
first choosing a convenient {\it fraction of small droplets}, which is then
kept (approximately) {\it fixed} throughout the run (typically we used $f_s
\approx 0.003$, that corresponds to removing $\sim 80$ droplets at each time 
step when the system contains $28000$ droplets).

Finally, we should take into account the shift of the droplets. For that one 
can use either the exact relation
\beq
{d\vec  r_i\over dt} =
2\sum_{j\ne i}{q_j\over |X_{i,j}|}{\vec X_{i,j}\over |X_{i,j}|},
\label{shift1}\eeq
where the charges $q_i=R_i\dot R_i$ are evaluated at each time 
step, or the heuristic formula
\beq
{d\vec r_i\over dt} =2\sum_{j\ne i}
\frac{K_{j,i}}{k_ik_j} 
\left( \frac{1}{R_i}-\frac{1}{R_j} \right)
{\vec X_{i,j}\over |X_{i,j}|^2}
\label{shift2}\eeq
discussed in the Introduction and Appendix A. Then we proceed according the 
following steps:
\begin{enumerate}
\item  
Pick $R_s$ such that the number of droplets with $R_n \leq R_s$ is
approximately $N_s=f_s N$. (In practice $R_s/\langle R\rangle$ is kept fixed -
in the scaling state this is the same as fixing $f_s$). Identify these as small
droplets and calculate all $V_m$ using Eq.(\ref{eq:new1b}).
\item 
An estimate $\tau^{(0)}$ for the time step is determined, as the interval 
during which all small droplets vanish; 
$$\tau^{(0)}=\max_m\{t_m\},$$  
where $t_m$ is determined by Eq.(\ref{tm}).
\item 
Calculate the velocities of the large droplets, $V_i=\sum_j v_{i,j}$ (summing
over neighbors of $i$ only - hence this step takes $\sim N$ operations). Note
that some droplet $i$ which has been classified as "large" may, in fact,
disappear during the interval $\tau^{(0)}$; this happens if the velocity is
such that $S_i<-V_i\tau^{(0)}$. All such droplets, if any, are collected and
reclassified as "small", a new value for $\tau$ is determined and the
velocities (both $V_m$ and $V_i$) are recalculated. Usually one such iteration
suffices to reach a self-consistent classification. 
\item 
Integrate Eq.(\ref{Final}) one step $\tau$; 
that is, calculate new areas according to Eq.(\ref{eq:new1aa}).
\item 
Calculate the shifts of the droplets according to either Eq.(\ref{shift1}) or 
Eq.(\ref{shift2}) and repeat the procedure.
\end{enumerate}
Since the relaxation time needed to reach the scaling state depends on how
close the initial configuration is to this state, one would like to choose the
initial configuration reasonably close to it. We prepared our initial state as
follows. We generated a set of $N=50000$ droplets with a distribution of radii
close to the expected one (determined from short preparatory runs). We
scattered them over the plane at random but so that each droplet is surrounded
by a small depletion zone (free of other droplets), to mimic the effect of
correlations that appear in the real evolving system. For this initial
configuration we started a rather large time step $\tau$ (and run the dynamics
without moving droplets) yielding only an approximation to the true dynamics.
This preliminary relaxation went on till the number of droplets was reduced to
$N=28000$. At this point a smaller time step was selected, the droplets were
allowed to move and we started to take measurements. All the results presented
below are averaged over $8$ runs in order to reduce statistical fluctuations.  

In our first simulations we did not take into account the shift of droplets. 
As is shown in Appendix A, this frozen droplets approximation might be good
even for fractions as large as $\varphi=0.13$. In order to test this, we
executed runs with frozen droplets and measured the fraction $f_c$ of droplets
crossing each other, as a function of the number of droplets in the system.
This data is presented in Fig. 1. It shows a considerable growth of $f_c$ with
time (from zero at $N=28000$ to $10\%$ at $N=1000$) and there is no tendency
towards stabilization. This proves that the droplets' motion has a considerable
effect on the dynamics of the system and, hence, the model of frozen droplets
is invalid for such large $\varphi$.

Taking the droplets' shift into account improves drastically this situation. 
We used both Eq.(\ref{shift1}) and Eq.(\ref{shift2}) (called models A and
B respectively; see below) for the droplets' motion.  Fig.1 shows that both models A and B
give rise to much smaller values of $f_c$ and that it saturates at long times. 
Interestingly, while there is no essential difference between these two models,
the heuristic model B exhibits slightly lower values of $f_c$ than model A. 
In order to test the convergence of the sum in Eq.(\ref{shift1}) we executed
runs with different numbers of terms in the sum taken into account; namely, we
varied the size of the summation box from $b=4.1\z$ to $b=5.8\z$ and found no 
noticeable difference in the behavior of $f$. Thus we conclude that we
have achieved proper convergence of Eq.(\ref{shift1}).

We now present results of our simulations, performed at the same relative area
fraction, $\varphi=0.13$, that was studied experimentally\cite{KS}. We measured
the position correlation function, $G(r)$; that is, the mean number of the
droplets whose centers lie within an annulus $[r,r+dr]$ around the center of a
given one. Fig 2. presents $G(r)$ obtained running model B at three consequent
moments of time, when the system contained 3000, 2000 and 1000 droplets
respectively (each of the curves is averaged over $8$ runs). We see that the
three curves coincide up to their fluctuations, which proves that the scaling
state has indeed been reached. The corresponding experimental data are also
presented in this figure; a good agreement is seen. Fig. 3 compares the
position correlation functions obtained by the models A and B (averaged over
time in the scaling state). This proves that our heuristic formula works
perfectly. The experimental $G(r)$ has a noticeable maximum at $r\approx 4.7$,
while for our curves the maximum is smaller; at the same time our curve is
rather close to the result of Masbaum (see Fig. 1a in ref.\cite{Masbaum}), that
has a small peak as well, which can not be distinguished clearly from the
fluctuations. Note that in ref.\cite{Masbaum} the data were taken with $N\sim
800$ droplets, whereas our data is well averaged (effectively it corresponds to
a system of about $50 000$ droplets in the scaling state), and the small peak
is definitely observed.

Analysis of the distribution of the droplets' radii at three consequent moments
of time, where the system contained 3000, 2000 and 1000 droplets respectively,
also shows that the distribution achieved its stationary form. No clear
difference between the models A and B was seen. The droplets' radii
distribution in the scaling state, averaged over time, as obtained from our
simulations (model B), is shown in Fig. 4 together with the experimental
results. A certain discrepancy is seen which, however, does not exceed by much
the statistical errors of the experimental points.

Finally, we measured the charge correlation functions\cite{KS}, that
contain more detailed information about the system, defined as follows. 
For a charge $q_i$ calculate $Q_+(r)$, the total amount of similar charge as
$q_i$ within an annulus $[r,r+dr]$ around $\vec r _i$ and define the function
$$g_+(r) = \langle q_i Q_+(r) \rangle.$$
Similarly we define $g_-(r)$ in terms of the {\it opposite} charges. These two
functions, as obtained by simulation of the two models, are presented in Fig. 5
together with the corresponding experimental data by Krichevsky and Stavans.
The agreement can be characterized as excellent. Again, there is no noticeable
difference between the results of models A and B.

One should notice that our definition of the charge correlation functions
$g_{\pm}(r)$, is not precisely the same as that of ref.\cite{KS}. Krichevsky
and Stavans smeared a droplet's charge on its perimeter before calculating
$Q_{\pm}(r)$, whereas we assigned a droplet's charge to it's center. We
believe, however, that the smearing of the charge on the droplet's perimeter is
no more than a way of smoothing the data and there is no essential difference
between the two definitions.

\section{Summary}
Ostwald ripening is the coarsening process, observed during the late stage of
the evolution of a two-phase system, where the droplets of the minority phase
exchange material by means of diffusion. This process leads towards a scaling 
state in which the characteristic length scale grows with time according to the 
scaling law $\bar R\sim t^{1/3}$, while all the statistical properties
(such as the droplets' size distribution, position correlation functions etc.), once rescaled, remains fixed.

The problem of calculating these characteristics has been studied in a number
of detailed numerical simulations, that take into account all the complicated
interactions between the droplets, mediated by the diffusion field. These
calculations, although being exact, do not contribute much to a qualitative
understanding of the importance of different components of the interaction
between the droplets. On the other hand, analytical mean field treatments
neglect all spatial effects and seem to be oversimplified.

The aim of this paper was to construct and test a "minimal extension" of the
mean field approach, that will take into account spatial effects, keeping only
the simplest interaction between the droplets. We calculated analytically an
approximate form of these interactions using a mean-field approach. Only
pairwise interactions between the droplets were preserved. We proposed a very
efficient numerical algorithm, which allows us to follow the evolution of tens
of thousands of droplets. We tested our approach by comparing its results with
the experimental data and found surprisingly good agreement at a relatively
large value of the minority phase area fraction, $\varphi=0.13$, where our
approach to the interaction between the droplets was not expected to work.

Trying to find the simplest model which reproduces the experimental data, we
examined the importance of a number of effects. Our findings are summarized as
follows:
\begin{enumerate}
\item 
Depletion zones have a considerable effect on the screening length, increasing
it from $\zeta=1.88\bar R$ to  $\zeta=2.73\bar R$. At the same time, as shown
in Table I, the presence of the depletion zones almost does not affect the
functional form of the pairwise interaction between the droplets.
\item
Our approach is based on the assumption of circular droplets and monopole +
dipole approximation for the diffusion field. The obtained agreement with the
experiment indicates that at the values of $\varphi$ studied higher multipoles
can be neglected.
\item
Even though inclusion of the depletion zones increases the screening length 
$\zeta$ too little to provide formal validity to our approximation, our 
simulations show that it works even for $\varphi=0.13$.
\item 
The effect of the droplets' motion is very important for large area fractions, 
although, formally, it could be regarded as adiabatically small compared to the droplets growth.
\item
The expression that determines the shift of the droplets requires summation over
a large number of droplets. We propose a much simpler heuristic formula
(containing a sum over the nearest neighbors only), that gives even better
results than the exact one.
\end{enumerate}
An advantage of our method is its computational efficiency; we are able to
choose time steps no smaller than required by physical reasonability,
eliminating a large number of droplets at each step. At the same time, the 
total area of the droplets is conserved exactly at each time step.
This makes our approach useful for extensive studies of the Ostwald problem.

\acknowledgments
This research was supported by grants from the Germany-Israel Science 
Foundation (GIF). B. L. thanks the Clore Foundation for financial support.
We thank O. Krichevsky, J. Stavans and D. Kandel for most useful discussions.
\newpage

\appendix
\section*{{\bf A~~~} Analysis of the formula for the droplets' shift.}
The droplets' shifts are determined by Eq.(\ref{shift}):
\beq
{d\vec  r_i\over dt} =
2\sum_{j\ne i}{q_j\over |X_{i,j}|}{\vec X_{i,j}\over |X_{i,j}|}.
\label{shifts}\eeq
First of all, using this formula we can show, that in the low area fraction
limit the shift of the positions can be neglected. Indeed, $\tau_{sh}$, the
characteristic time of a shift of a droplet's center of mass is given by
\beq
\tau_{sh}^{-1}\sim {1\over X_{i,j}}\left|{d\vec  X_{i,j}\over dt}\right|
\sim {q_i\over X_{i,j}}.
\label{tsh}\eeq
The characteristic growth time of the droplet is (see Eq.(\ref{eq:dR}))
\beq
\tau_{gr}^{-1}\sim {1\over R_i}\left|{d\vec R_i\over dt}\right|
\sim {q_i\over R_i^2}.
\label{tgr}\eeq
Comparing Eqs.(\ref{tsh}) and (\ref{tgr}) we see that 
\beq
\tau_{sh}^{-1}\sim\tau_{gr}^{-1}{R_i^2\over X_{i,j}^2}=
\tau_{gr}^{-1}{\varphi\over\pi}.
\label{tau}\eeq
That is, for small area fractions the motion of the droplets' centers is
adiabatically slower than their growth. Consequently, one can neglect the
droplets' motion and the system is characterized only by the dynamics of the
droplets' radii as determined by Eq.(\ref{eq:dRdt})- (\ref{eq:Lij}). Formally,
one can expect this approximation to be valid even for $\varphi=0.13$ (used in
the experiments by Stavans and Krichevsky) and we have tried it in our work
(see Section IV and Fig. 1). Our simulations have shown, however, that 
neglecting the droplets' motion gives wrong results and, therefore, the 
dynamics of the $\vec r_i$ has been taken into account.

Secondly, note that a rough estimate of the sum on the r.h. of
Eq.(\ref{shifts}) indicates that it exhibits bad convergence properties. 
Assuming $q_j$ to be uncorrelated random variables with zero mean we get
\beq
\left\langle\left({d\vec  r_i\over dt}\right)^2\right\rangle \sim
 \langle q^2\rangle\sum {1\over |X_{i,j}|^2}\sim \langle q^2\rangle\log L_s,
\eeq
where $L_s$ is the size of the system. A more accurate treatment implies,
however, that the sum does converge. According to
Eqs.(\ref{eq:dR})-(\ref{eq:Linv})
$$q_j= R_j{dR_j\over dt}=
\sum_{m (\neq j)}K_{j,m}\Delta_{j,m},$$
where 
$$\Delta_{j,m}=\frac{1}{k_mk_j} \left( \frac{1}{R_m}-\frac{1}{R_j} \right)$$
and we use the abbreviations 
$K_{j,i}=K_0(X_{j,i}/\z ),~k_j=K_0(R_j/\z)$. 
Then Eq.(\ref{shifts}) becomes:
\beq
(1/2){d\vec X_i\over dt}=
\sum_{j~(\ne i)}Q_j^{(i)}{\vec X_{i,j}\over |X_{i,j}|^2} +
\sum_{j~(\neq i)}\sum_{m(~\neq j,i)}
Q_j^{(m)}{\vec X_{i,j}\over |X_{i,j}|^2},
\label{terms}\eeq
where
$$Q_i^{(j)}\equiv K_{j,i}\Delta_{j,i}$$
denotes the part of the charge on the $i$-th droplet that is induced by the
$j$-th one.  The first term of Eq.(\ref{terms}) determines the shift of the
droplet due to the direct material transfered between this droplet and its
$j$-th neighbors. This sum converges very well because $Q_i^{(j)}$ decreases
exponentially with distance $X_{i,j}$. The second term accounts for
the effect of the redistribution of the material between the $j$-th and $m$-th
droplets. Since the internal sum (on $m$) contains a short-range factor
$K_{j,m}$, it is actually over a $\zeta\times\zeta$-box around the $j$-th
droplet. In this double sum, each term 
$$Q_j^{(m)}{\vec X_{i,j}\over |X_{i,j}|^2}$$
can be paired with:
$$Q_m^{(j)}{\vec X_{i,m}\over |X_{i,m}|^2}.$$
Since $Q_j^{(m)}=-Q_m^{(j)}$, these two contributions can be considered as a
$dipole$. Thus, each $\zeta\times\zeta$-box represents a dipole $\vec P$
(randomly directed) with the dipole moment $P\sim q\zeta$, where $q$ is the
characteristic scale of $Q_m^{(j)}$. Thus, the second term in Eq.(\ref{terms})
is now estimated as:
\beq
{\rm second~term}\sim
\sum_{\zeta\times\zeta-{\rm boxes}}
{\vec P\over |X_{i,j}|^2}-{2(\vec P\cdot\vec X_{i,j})\vec X_{i,j}
\over |X_{i,j}|^4}\sim\sum {\vec P\over |X_{i,j}|^2}.
\eeq
The mean of this expression is zero while the mean square deviation is given by
\beq
\sim q^2\zeta^2\sum {1\over |X_{i,j}|^4},
\eeq
which converges to a finite value. Thus, when calculating the sum in 
Eq.(\ref{shifts}),
we can restrict ourselves to only several nearest layers of neighbors. 

Finally, one can use an even simpler heuristic formula for calculating the
shift of the droplets. The meaning of Eq.(\ref{terms}) is that the motion of
$i$-th droplet has two sources. The first is the material transfered between
this droplet and its $j$-th neighbors (the first term of Eq.(\ref{terms})). The
second is due to redistribution of the material between the surrounding $j$-th
droplets them selves (the second term). Although these contributions are of the
same order, in our case we have a reason to drop the second term (although it
does not simplify computations, it does makes the model physically simpler). 
The shift of the droplets has a noticeable effect only at relatively large
fractions, where the interaction between the next nearest neighbors is
considerably suppressed. Then, for a fixed configuration of the nearest
neighbors of the $i$-th droplet we can vary the configuration of its next
nearest neighbors. This manipulation will not affect the first term of the
Eq.(\ref{terms}), while it will reduce its second term. Thus, in the mean field
spirit of our model, we can average the shift velocity of the $i$-th droplets
over various configurations of its next nearest neighbors. Thus, we finally get:
\beq
{d\vec X_i\over dt} =2\sum_{j\ne i}
\frac{K_{j,i}}{k_ik_j} 
\left( \frac{1}{R_i}-\frac{1}{R_j} \right)
{\vec X_{i,j}\over |X_{i,j}|^2}.
\eeq

Although the approximation leading to this formula is not based on a rigorous
expansion, our simulations show that it works as well as the more rigorous
Eq.(\ref{shifts}). At the same time it is much more economic because it
requires the summation only over the nearest neighbors.
\newpage

\section*{{\bf B~~~} Depletion zones} 
The simplest possible improvement over the mean-field approximation can be
obtained by including some of the correlations between the positions of the
droplets' centers.
In this Appendix we will take into account the simplest manifestation of these
correlations $-$ the so called {\it depletion zones}. These are the regions
around each of the droplets, from which all possible neighbors are excluded (by
geometrical steric constraints - the distance between the centers of any two
droplets must exceed at least the sum of their radii). 

The $exact$ equation for the diagonal elements of the matrix $\hat\phi$ 
is given by Eq.(\ref{eq:ex}),
whereas for the off-diagonal elements it is Eq.(\ref{eq:expl})
\beq
\sum_{j \neq i,k}{1\over \log R_j/R_0} \log (X_{i,j}/R_0) \phi_{j,k}+
\phi_{i,k}= -\gamma_k\log X_{i,k}/R_0.
\label{ex}\eeq
This 
has been replaced in our mean field approximation by the following integral 
equation for the smooth function $\phi(x)$ (see Eq.(\ref{eq:naiv})):
\beq
-{1\over 2\pi}\zeta_{0}^{-2} \int\log (X_{i,j}/R_0)\phi(X_{j,k})d^2r_j+
\phi(X_{i,k})=-\gamma \log X_{i,k}/R_0.
\label{naiv}\eeq
Clearly, by using the integral of Eq.(\ref{naiv}) we implicitly assume that the
distribution of ${\vec r}_j$, the positions of the centers of droplets $j$, are
independent of their distance from the one at ${\vec r}_i$.  This homogeneity
assumption may serve as a reasonable approximation as long as $\bar X$, the
mean distance between neighbors, is much greater than $d$, the typical radius
of the depletion zones.  When, however, the density increases to the extent
that ${\bar X} \sim d$, the inhomogeneity of the distribution of the
$j$-droplets around the fixed droplets $i$ and $k$ can no longer be ignored.
One should emphasize that other effects, such as correlations between different
"charges" and between the droplets' sizes may also be of importance and can
also possibly affect the elements of the inverse matrix we are calculating.
Nevertheless, here we take into account only the correlations that ensure that
the areas of two neighbors do not overlap. In our actual numerical calculations
even this is done only approximately, by introducing a {\it uniform} sized
depletion zone, neglecting its fluctuations as well as dependence on the
droplets' radii.  

Correlation between the positions of the droplets can be incorporated in our
approximate treatment by replacing the sum in Eq.(\ref{ex}) by the following
integral operator:
\beq
\hat S_{int}\phi=  -{1\over 2\pi}\zeta_{0}^{-2}
\int \log (X_{i,j}/R_0) P(\vec r_i,\vec r_j,\vec r_k) \phi(X_{j,k})d^2r_j, 
\label{eq:depz1}\eeq 
where $P$ is the probability of finding a droplet centered at the point $\vec
r=\vec r_j$, given that there are droplets at $\vec r=\vec r_i$ and $\vec
r=\vec r_k$.  We approximate this probability (three-point correlation
function) by representing it as the product of pair correlation functions:
\beq
P(\vec r_i,\vec r_j,\vec r_k)=g(X_{i,j})g(X_{j,k}),
\label{eq:depz2}
\eeq
where $g(X_{i,j})$ is the probability of finding a droplet $j$ at distance
$X_{i,j}$ from the center of droplet $i$; this function is normalized such 
that $g(r)\rightarrow 1$ at $r\rightarrow\infty$. Within this approximation 
Eq.(\ref{ex}) is replaced by
\beq
\hat S_{int}\phi+
\phi(X_{i,k})=-\left(1+{\phi_{k,k}\over\log (R_k/R_0)}\right)
\log X_{i,k}/R_0
\label{eq:depz3}\eeq
where now we have
\beq
\hat S_{int}\phi=  -{1\over 2\pi}\zeta_{0}^{-2}
\int\log (X_{i,j}/R_0)g(X_{i,j})g(X_{j,k})\phi(X_{j,k})d^2r_j.
\label{eq:depz4}
\eeq
In order to solve equation (\ref{eq:depz3}) we first try to bring it as close
to the form of Eq.(\ref{naiv}) as we can, and then use the same method of
solution as was used there. To this end we first rewrite the expression for 
${\hat S}_{int}$ as follows:
\begin{eqnarray}
\hat S_{int}\phi =  &-&{1\over 2\pi}\zeta_{0}^{-2}
\int\log (X_{i,j}/R_0)g(X_{i,j})\phi(X_{j,k})d^2r_j \nonumber \\
&+& {1\over 2\pi}\zeta_{0}^{-2}
\int\log (X_{i,j}/R_0)g(X_{i,j})[1-g(X_{j,k})]\phi(X_{j,k})d^2r_j.
\label{eq:depzSint1}
\end{eqnarray}
The pair correlation function $g(X)$ vanishes for short distances, $X < d$, and
$g \approx 1$ for large $X$. Therefore, we get non-vanishing contributions to
the second term in Eq.(\ref{eq:depzSint1}) only when  $X_{j,k}<d$.  For
$X_{i,k} \gg d$, we have in this region $g(X_{i,j}) \approx 1$ and $\log
(X_{i,j}/R_0) \approx \log (X_{i,k}/R_0)$, so that 
\begin{eqnarray}
\hat S_{int}\phi &\approx&
-{1\over 2\pi}\zeta_{0}^{-2}
\int\log (X_{i,j}/R_0)g(X_{i,j})\phi(X_{j,k})d^2r_j \nonumber \\
&-&{1\over 2\pi}\zeta_{0}^{-2}\log (X_{i,k}/R_0)
\int[1-g(X_{j,k})]\phi(X_{j,k})d^2r_j.
\label{eq:depzint1}
\end{eqnarray}
The last formal step we take is to express the first integral here as the sum
of two terms, using $-g=[1-g]-1$, which leads to our final expression for $\hat
S_{int}\phi$:
\begin{eqnarray}
 \hat S_{int}\phi = &-&{1\over 2\pi}\zeta_{0}^{-2}
\int \log (X_{i,j}/R_0) \phi(X_{j,k})d^2r_j \nonumber \\
&+& {1\over 2\pi}\zeta_{0}^{-2}
\int \log (X_{i,j}/R_0)\left[1- g(X_{i,j})\right] \phi(X_{j,k})d^2r_j
\nonumber \\
&-&{1\over 2\pi}\zeta_{0}^{-2}\log (X_{i,k}/R_0)
\int[1-g(X_{j,k})]\phi(X_{j,k})d^2r_j.
\label{eq:depzSint}
\end{eqnarray}
Our basic equation (\ref{eq:depz3}) takes now the form
\begin{eqnarray}
&-&{1\over 2\pi}\zeta_{0}^{-2}
\int \log (X_{i,j}/R_0) \phi(X_{j,k})d^2r_j +\phi(X_{i,k})
+{1\over 2\pi}\zeta_{0}^{-2}
\int F(X_{i,j}) \phi(X_{j,k})d^2r_j \nonumber \\
&=&-\gamma_k\log(X_{i,k}/R_0),
\label{eq:depznaiv} 
\end{eqnarray}
where
\beq
\gamma_k=1+{\phi_{k,k}\over\log (R_0/R_k)}+{1\over 2\pi}\zeta_{0}^{-2}
\int(1-g(X_{j,k}))\phi(X_{j,k})d^2r_j
\label{eq:depzgam}\eeq
and
\beq
F(X)= \log (X/R_0) [1-g(X)].
\label{eq:depzFX}\eeq
Note that equation (\ref{eq:depznaiv}) is very close in form to
Eq.(\ref{naiv}); the only significant difference being the appearance of a
new term - the integral over the function $F$, which contains explicitly the
correlation function $g$. Clearly when $g(X)\equiv 1$ we recover
Eq.(\ref{naiv}).  As was the case there, the
solution of Eq.(\ref{eq:depznaiv}) is symmetric if and only if $\gamma_k$ does
not depend on $k$, $\gamma_k \approx \gamma$, which  is achieved by imposing
$\phi_{k,k}=0$. This leads again to a condition that determines the value of
the parameter $R_0$ (see Eq.(\ref{eq:depzR0}) below). Although 
now $\gamma\ne 1$, (see Eq.(\ref{eq:depzgam})), its value is
unimportant for the simulations since it can be absorbed in the definition of
the time scale in the dynamic equation. Hence below we will set $\gamma =1$.  

We solve Eq.(\ref{eq:depznaiv}) by first applying the $\nabla^2_i$ operator to
the equation, which yields
\beq
-\zeta_0^{-2}\phi(X_{i,k})+\nabla^2_i\phi(X_{i,k})+\frac{1}{2\pi} \zeta_0^{-2}
\nabla^2_i \int F(X_{i,j}) \phi(X_{j,k})d^2r_j
=-2\pi \gamma \delta(X_{i,k}).
\label{eq:depzdif1}\eeq
Denote the Fourier transform of $F(X)$ by
\begin{eqnarray}
\tilde F(q)&=&\frac{1}{2\pi} \int d^2 r \log \left(\frac{r}{R_0}\right)
[1-g(r)] e^{i{\vec q} \cdot {\vec r}} \nonumber \\
&=& \int_0^\infty rdr  \log \left(\frac{r}{R_0}\right) [1-g(r)] J_0(qr).
\label{eq:depzFT}
\end{eqnarray}
Note that the integral converges, since $g(r) \rightarrow 0$ fast for large 
$r$. Upon Fourier transforming Eq. (\ref{eq:depzdif1}) becomes 
\beq
-\zeta_0^{-2}\tilde\phi(q)-q^2\tilde\phi(q)- \zeta_0^{-2}q^2 
\tilde F(q)\tilde\phi(q) = -2\pi \gamma 
\label{eq:depzdif2}
\eeq
yielding the solution for $\tilde\phi(q)$ (from now on we set $\gamma=1$)
\beq
\tilde\phi(q)= \frac{2\pi}{q^2} \frac{1}{1+\zeta_0^{-2}[\tilde F(q)+q^{-2}]}.
\label{eq:depzphiq}
\eeq
Once $\tilde\phi(q)$ has been evaluated, the function $\phi(X)$ is obtained
by the inverse Fourier transform. 
>From this point on we work with the specific (step-function) form of $g(X)$:
\begin{eqnarray} 
g(X)= \cases{ 0  & \qquad $X < d$  \cr
              1  & \qquad $X > d$ \cr}
\label{eq:depzg}
\end{eqnarray}
For this form of $g$ the integral in Eq. (\ref{eq:depzFT}) can be calculated
analytically, integrating by parts and using the identities $[xJ_1(x)]'=
xJ_0(x)$ and $~~[J_0(x)]'=J_1(x)$, to get
\beq
\tilde F(q)=q^{-2} \{ - qdJ_1(qd)\log(R_0/d) +\left[1-J_0(qd)\right] \}.
\label{eq:depzFq}\eeq
We now substitute this in (\ref{eq:depzphiq}) and perform the angular
integration in the inverse Fourier transform, to arrive at  
the following expression for $\phi (r)$:
\beq
\phi(r)=\int_0^{\infty}{J_0(qr)\over q^2+\zeta_0^{-2}-[\zeta_0^{-2}
\log (R_0/d)]qdJ_1(qd)+\zeta_0^{-2}\left[1-J_0(qd)\right]}qdq.
\label{eq:depzphir}\eeq 
We cannot perform this integral analytically, but it is simple to evaluate 
by numerical integration. Once this has been done, the parameter $R_0$ can be
found by imposing the condition $\phi_{k,k}=0$, which takes the form (analogous
to Eq.(\ref{eq:diag00}))
\beq
\phi_{k,k} = \frac{1}{2\pi} \zeta_0^{-2} \int \log \left( \frac{r}{R_0} \right)
g(r) \phi(r) d^2r =0
\label{eq:depzdiag00}
\eeq
which becomes the following equation for $R_0$:
\beq
\log R_0={\int_d^{\infty}\log(r)\phi(r)rdr\over\int_d^{\infty}\phi(r)rdr}
\label{eq:depzR0}
\eeq
Once $R_0$ has been determined we can evaluate the parameter $\zeta_0$, 
using its definition
\beq 
\zeta_0^{-2}=2\pi n\langle{1\over \log(R_0/R_i)}\rangle
\label{eq:depzz0}
\eeq
For a given droplet configuration one can now proceed to find $\phi(r)$ by
taking the following steps. First, determine the radius of the depletion zone
$d$ from the measured correlation function of the droplets (such as Fig. 3). 
Then solve Eqs.(\ref{eq:depzphir},\ref{eq:depzR0}) and (\ref{eq:depzz0}). This
can be done iteratively by initializing the procedure using the results of the
correlation-free theory for $\phi,\zeta_0$ and $R_0$. The next iterate of
$\phi(r)$ is evaluated using Eq.(\ref{eq:depzphir}); this new $\phi (r)$ is
then used in Eq.(\ref{eq:depzR0}) to yield the new value of $R_0$; this, in
turn, is used in Eq.(\ref{eq:depzz0}), together with the distribution of the
droplets' radii, to yield the new iterate of $\zeta_0$. The procedure is then
repeated until convergence is reached. In practice we used a few simplifying
steps, which made computation faster without significantly altering the result.
The first simplification consists of setting
\beq
d =\kappa \langle R \rangle
\label{eq:depzdd}\eeq
for an entire run, instead of determining it from the correlation function at 
each time step. We found that values in the range $2.0 \leq \kappa \leq 2.8$ 
fit the correlation function quite well. The numerical results were obtained
using the fixed value $\kappa = 2.15$. A second time-saving simplification is
to use the approximation
\beq
\zeta_0^{-2}=2\pi n{1\over \log(R_0/\langle R \rangle)}
\label{eq:depzz0a}\eeq
to calculate $\zeta_0$, instead of evaluating the average $\langle
1/\log(R_0/R_i)\rangle$ over the measured distribution of droplet sizes. Recall
also that the density of droplets, $n$, is related to their relative area
fraction, $\varphi$, by $n=\varphi/\pi \langle R^2 \rangle$. In the spirit of
the previous approximation we replace $\langle R^2 \rangle \approx \langle R
\rangle^2$, so that Eq.(\ref{eq:depzz0a}) becomes
\beq 
\zeta_0^{-2}={2 \varphi \over
\langle R \rangle^2}{1 \over \log(R_0/\langle R \rangle)}
\label{eq:depzz0app}\eeq 
Using Eqs.(\ref{eq:depzdd}) and (\ref{eq:depzz0app}) simplifies our numerical
scheme significantly. To demonstrate the effect of including the depletion
zones in the calculation we present in Table I the results of calculations
performed in the scaling regime at four different values of $\varphi$. The
various quantities presented were obtained as follows: $R_0$ and $\zeta_0$ are
(simultaneous) solutions of Eqs. (\ref{eq:depzphir}), (\ref{eq:depzR0}) (with
$d=2.15 \langle R \rangle$ in both) and (\ref{eq:depzz0app}). $\zeta_0^{mf}$ is
the value of the screening length as obtained by setting $d=0$. The screening 
length $\zeta_{scr}$ is the first moment of the function $\phi(r)$,
\beq
\zeta_{scr}^{2}=\int_d^{\infty} \phi (r) rdr
\label{depzmoment}\eeq
calculated by numerical integration of Eq.(\ref{eq:depzphir}).

As we see, for very small fractions the screening length is less than
$\zeta_0^{mf}$. This is surprising, since we expected inclusion of the
depletion zones to increase the screening regions.  For the larger fractions,
however, we indeed see that the effect of including the depletion zone changes
sign: the renormalized screening length becomes greater than the mean field
result, as was expected.  

Another important observation we should make is that as $\varphi$ increases,
$\zeta$ decreases and becomes comparable to the average radius (in units of
which it is given in Table I). At such $\varphi$, $\zeta$ obviously cannot be
interpreted as a "screening length", because $\zeta\approx\langle R\rangle$
means that there are no droplets in the "screening zone". 

Another question that we studied in detail concerns the extent to which
introducing the depletion zone affects the function $\phi(r)$. Clearly, when we
are not in the limit of very small $\varphi$ it is no longer given by the
MacDonald function. Table II contains $\phi(r)$, as obtained at $\varphi= 0.13$ 
by the procedure outlined above: setting $d=2.15\langle R\rangle$, and
simultaneously solving Eqs.(\ref{eq:depzphir},\ref{eq:depzR0}) and
(\ref{eq:depzz0app}). Comparing $\phi(r)$ with the simple mean-field result
$K_0(r/\zeta_0^{mf})$ reveals that the renormalization due to inclusion of the
depletion zones leads to significant change of the matrix elements $\phi(r)$. 
On the other hand, comparing $\phi(r)$ with the function $K_0(r/\zeta_0)$ we
see that $\phi(r)$ does not differ much from MacDonald's function, provided the
$renormalized$ $\zeta_0$ is used. Therefore in principle this function can be
used in calculations as a fairly good approximation for $\phi(r)$. The
numerical results presented in Section IV were obtained using $\zeta_0=2.73$,
according to Table I.  

\section*{{\bf C~~~} Derivation of the condition for $R_0$.}
The condition (45) on $R_0$:
\beq
{1\over 2\pi}\zeta_{0}^{-2}
\int\log (X_{k,j}/R_0)K_0(X_{j,k}/\zeta_0)d^2r_j= 0
\label{A:diag}\eeq
can be simplified by representing 
\[
\log(X_{i,j}/R_0)=\log(X_{i,j}/\zeta_0)+\log(\zeta_0/R_0)
\]
so that Eq. (\ref{A:diag}) becomes
\beq
B+A\log(\zeta_0/R_0)=0
\label{eq:B+A}\eeq
where the constant $A$ has the value
\beq
A={1\over 2\pi}\zeta_{0}^{-2}\int
K_0(X_{j,k}/\zeta_0)d^2r_j=1
\label{eq:A}\eeq
because of the normalization of $K_0(x)$, while the other constant, $B$, is
given by
\beq
B={1\over 2\pi}\zeta_{0}^{-2}\int\log (X_{k,j}/\zeta_0)
K_0(X_{j,k}/\zeta_0)d^2r_j
\label{eq:B}\eeq
In order to evaluate $B$ we note that the solution of Eq.(\ref{eq:naiv}) is
given, for any value of $R_0$, by Eq.(\ref{eq:phiik}). Therefore we can
substitute $\phi(X_{i,k})$ from Eq.(\ref{eq:phiik}) in Eq.(\ref{eq:naiv}) and
choose for $R_0$ the special value $R_0 = \zeta_0$, to get the identity
\beq
-{1\over 2\pi}\zeta_{0}^{-2}\int 
\log(|\vec r-\vec r'|/\zeta_{0}) K_0(r'/\zeta_{0})d^2r'+
K_0(r/\zeta_{0})=-\log (r/\zeta_{0})
\label{eq:iden1}\eeq
It is trivial to see that in the limit ${\vec r} \rightarrow 0$ the integral on
the l.h.s. of this identity becomes precisely equal to $B$ (as given by Eq. 
(\ref{eq:B})). For ${\vec r} \rightarrow 0$ we can use the small $r$ limit
(\ref{eq:zero}) of $K_0$ so that for very small $r$ our identity becomes
\[
-B -\log\left(\frac{r}{2 \zeta_0} \right) - C = -\log
\left(\frac{r}{\zeta_0} \right)
\]
so that we find 
\beq
B=\log2 - C
\label{eq:Bvalue}
\eeq
($C$ is Euler's constant). Using the values of $A$ and $B$ in Eq.(\ref{eq:B+A}) 
it becomes 
\beq
\log 2 - C + \log (\zeta_0 / R_0 ) = 0  
\label{eq:con1}
\eeq
providing a new connection between $R_0$ and $\zeta_0$. This is to be used
together with Eq.(\ref{eq:z}) to solve for both $R_0$ and $\zeta_0$. Since
Eq.(\ref{eq:z}) contains $\log (R_0/\langle R \rangle )$, it is convenient to
add and subtract $\log\langle R\rangle $ from Eq.(\ref{eq:con1}) and to 
rewrite it as
\beq
\log(R_0/<R>)=\log (2 \zeta_0/\langle R \rangle ) - C
\label{eq:con2}\eeq
If we have $\langle R\rangle\ll \zeta_0$ the right hand side of
Eq.(\ref{eq:con2}) is $\approx K_0(\langle R \rangle /\zeta_0)$, so that
(\ref{eq:con1}) becomes $\log(R_0/<R>)\approx K_0(<R>/\zeta_0)$. Finally, we
can write, with the same accuracy:
\beq
\left\langle {1\over\log (R_0/R)}\right\rangle 
\approx \left\langle {1\over K_0( R /\zeta_0)}\right\rangle.
\label{eq:con0}\eeq

\section*{{\bf D~~~} The small parameter of the approximation.}
In the mean field limit, when $R_0\sim \zeta_0 \gg R_i$ the quantity
$1/\log(R_0/R_i)$ does not vary much and one can take it out of the sum and
replace it by its mean value. Secondly, we see that there is a scale $\zeta_0$
such that $\phi$ does not change much over distances much less than $\zeta_0$.
Let us divide the plane into boxes $b$ of size $\zeta_0\times\zeta_0$. Then the
sum that still remains can be rewritten as 
\beq
\sum_{j \neq i,k}\log (X_{i,j}/R_0) \phi_{j,k}=
\sum_{b}~~~\sum_{j \in b , (j\neq i,k)}\log (X_{i,j}/R_0) \phi_{j,k}
\eeq
Actually, only the  few boxes located near $\vec r_k$ have a significant
contribution to the sum; the contribution of all the others is exponentially
small due to the screening effect.

The expression being summed does not change much inside each box and therefore,
if the mean number of the droplets in the box, $N_{\zeta}$, is large enough,
the internal sum within each box can be replaced by the integral. In this
case, according to the theorem of large numbers
\beq
\left|\sum_{j \in b}\log (X_{i,j}/R_0) \phi_{j,k}- 
n\int\log (X_{i,j}/R_0) 
\phi(X_{j,k})d^2r_j\right|\sim {1\over\sqrt{N_{\zeta}}}.
\eeq
This gives rise to the following condition for the validity of our 
approximation: 
\beq
N_{\zeta}=n\zeta_0^2\gg 1
\label{eq:n}\eeq
But Eq.(\ref{eq:z}) implies that 
$n\zeta_{0}^2\sim(2\pi)^{-1}\log(R_{0}/\langle R\rangle)$ 
and, as we have shown $\zeta_0 \approx R_0$, so that 
\beq
n\zeta_{0}^2\sim\log(\zeta_{0}/\langle R \rangle )
\label{eq:nz1} \eeq
On the other hand, multiplying Eq.(\ref{eq:z}) by $\langle R \rangle ^2$ and
using the definition of the area fraction $\varphi=\pi n\langle R^2\rangle\sim
\pi n\langle R\rangle ^2$ yields $\log (\zeta_0 / R_0)\sim\log(\varphi^{-1})$.  This, together with Eqs.(\ref{eq:n}-\ref{eq:nz1}) means that 
$$\log (\varphi^{-1}) \gg 2\pi.$$
\newpage

\vbox{
\begin{table}
\vskip0.8cm 
\hskip3.0cm
\begin{tabular}{llllll}\hline
\vline~$~~~~~\varphi~~~~~$ &\vline ~~~$R_0~~~$ &\vline ~~~$\zeta_0^{mf}~~~$ &\vline ~~~$\zeta_0$~~~&\vline ~~~$\zeta_{scr}~~~$
&\vline\\ \hline 
\vline ~0.001 &\vline ~51.35 &\vline ~45.99&\vline ~~45.96&\vline ~~45.922
&\vline\\ \hline
\vline ~0.01  &\vline ~15.59&\vline  ~11.77 &\vline ~~12.13&\vline ~~11.71
&\vline  \\ \hline
\vline ~0.05 &\vline ~8.10&\vline  ~4.08&\vline ~~4.73&\vline ~~4.00
&\vline \\ \hline
\vline ~0.13 &\vline ~6.14&\vline ~1.88 &\vline ~~2.73&\vline ~~1.94
&\vline
\end{tabular}
\vskip2pc
\caption{
Various quantities related to the self consistent determination of $\phi(r)$ 
(with the depletion zones taken into account) for different area fraction 
$\varphi$. $R_0$ and $\zeta_0$ are (simultaneous) solutions of Eqs. 
(\protect\ref{eq:depzphir},\protect\ref{eq:depzR0}) (with $d=2.15\langle
R\rangle$ in both) and (\protect\ref{eq:depzz0app}); $\zeta_{mf}$ is the value
of the screening length as obtained in the Section III (neglecting the effect
of the depletion zones); $\zeta_{scr}$ is determined by Eq.
(\protect\ref{depzmoment}) with $\phi(r)$ obtained by numerical integration. 
All lengths are given in units of $\langle R \rangle$.
}
\end{table}
}

\vbox{
\begin{table}
\begin{tabular}{lllll}
\vline ~~~r/x~~~ &\vline  ~~~$\phi(r)$~~~  &\vline  ~~~$K_0(r/\zeta_0)$~~~ &\vline ~~~$K_0(r/\zeta_{mf})$~~~ &\vline\\ \hline
\vline ~~~  .722  &\vline ~~~ 2.884  &\vline ~~~ 2.617 &\vline ~~~  1.205  &\vline \\ \hline
\vline ~~~  .884 &\vline ~~~  1.914  &\vline ~~~ 1.773  &\vline ~~~ .711  &\vline \\ \hline
\vline ~~~ 1.045  &\vline ~~~ 1.320 &\vline ~~~  1.218  &\vline ~~~ .425  &\vline \\ \hline
\vline ~~~ 1.206  &\vline ~~~ .901  &\vline ~~~ .845  &\vline ~~~ .257  &\vline \\ \hline
\vline ~~~ 1.420  &\vline ~~~ .546  &\vline ~~~ .525  &\vline ~~~ .133  &\vline \\ \hline
\vline ~~~ 1.689  &\vline ~~~ .297  &\vline ~~~ .294  &\vline ~~~ .059  &\vline \\ \hline
\vline ~~~ 1.904  &\vline ~~~ .185  &\vline ~~~ .186  &\vline ~~~ .031  &\vline \\ \hline
\vline ~~~ 2.012  &\vline ~~~ .147  &\vline ~~~ .148  &\vline ~~~ .022  &\vline \\ \hline
\vline ~~~ 2.118  &\vline ~~~ .118  &\vline ~~~ .118  &\vline ~~~ .016  &\vline \\ \hline
\vline ~~~ 2.333  &\vline ~~~ .075  &\vline ~~~ .076  &\vline ~~~ .008  &\vline \\ \hline
\vline ~~~ 2.548  &\vline ~~~ .046  &\vline ~~~ .048  &\vline ~~~ .004  &\vline \\ \hline
\vline ~~~ 2.763  &\vline ~~~ .027  &\vline ~~~ .031  &\vline ~~~ .002  &\vline \\ \hline
\vline ~~~ 2.97  &\vline ~~~ .016  &\vline ~~~ .020  &\vline ~~~ .001  &\vline \\ \hline
\vline ~~~ 3.031  &\vline ~~~ .014  &\vline ~~~ .018  &\vline ~~~ .001  &\vline \\ \hline
\vline ~~~ 3.353 &\vline ~~~ .0087 &\vline ~~~ .0095 &\vline ~~~ .00047 &\vline 
\end{tabular}
\vskip2pc
\caption{ 
Interaction function $\phi(r)$ for the case of depletion zones calculated at
the area fraction $\varphi=0.13$. It is compared with the MacDonald's function
with the renormalized screening parameter $\zeta_0$ (third column), and with
the unrenormalized, mean-field screening length $\zeta_0^{mf}$ (fourth column).
Note that the distance $r$ is measured in units of the mean distance between 
neighbor droplets, $x=1/\protect\sqrt{n}$.
}
\end{table}
}
\input epsf


\begin{figure}[h]
\epsfxsize=140mm
        \centerline{ \epsffile{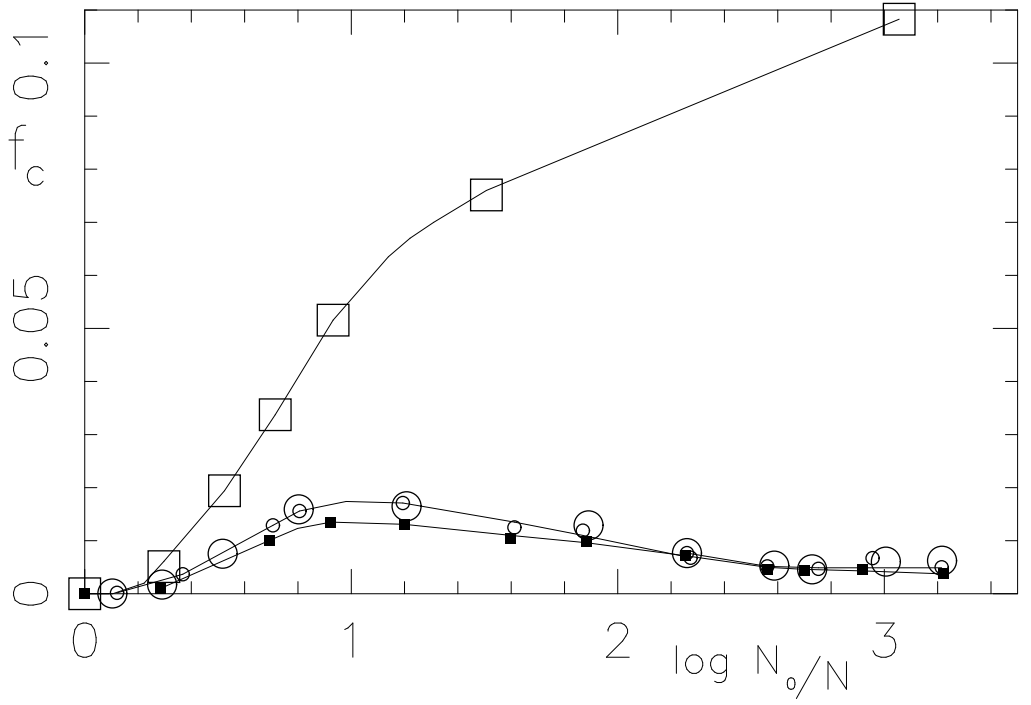}  }           
    \vspace{5mm}   
\caption{
Time dependence of $f_c$, the fraction of crossing droplets, as obtained from
the frozen droplets model ({\it empty squares}); from model B, that uses the
heuristic Eq.(\protect\ref{shift2}) for the droplets' shift ({\it full
squares}) and from model A, that uses the exact Eq.(\protect\ref{shift1}) for
the droplets shift ({\it large circles} and {\it small circles} for small and
large summation block sizes: $b=4.1\zeta$ and $b=5.8\zeta$ respectively; see
explanations in text). The lines are guides for eye. The growth of $f_c$
observed for the frozen droplets model indicates that it is not valid for the
area fraction $\varphi=0.13$. On the other hand, once the droplets are allowed
to move, $f_c$ decreases to very small values irrespective of whether model B
or A is used.
}
\end{figure}

\begin{figure}[h]
\epsfxsize=140mm
        \centerline{ \epsffile{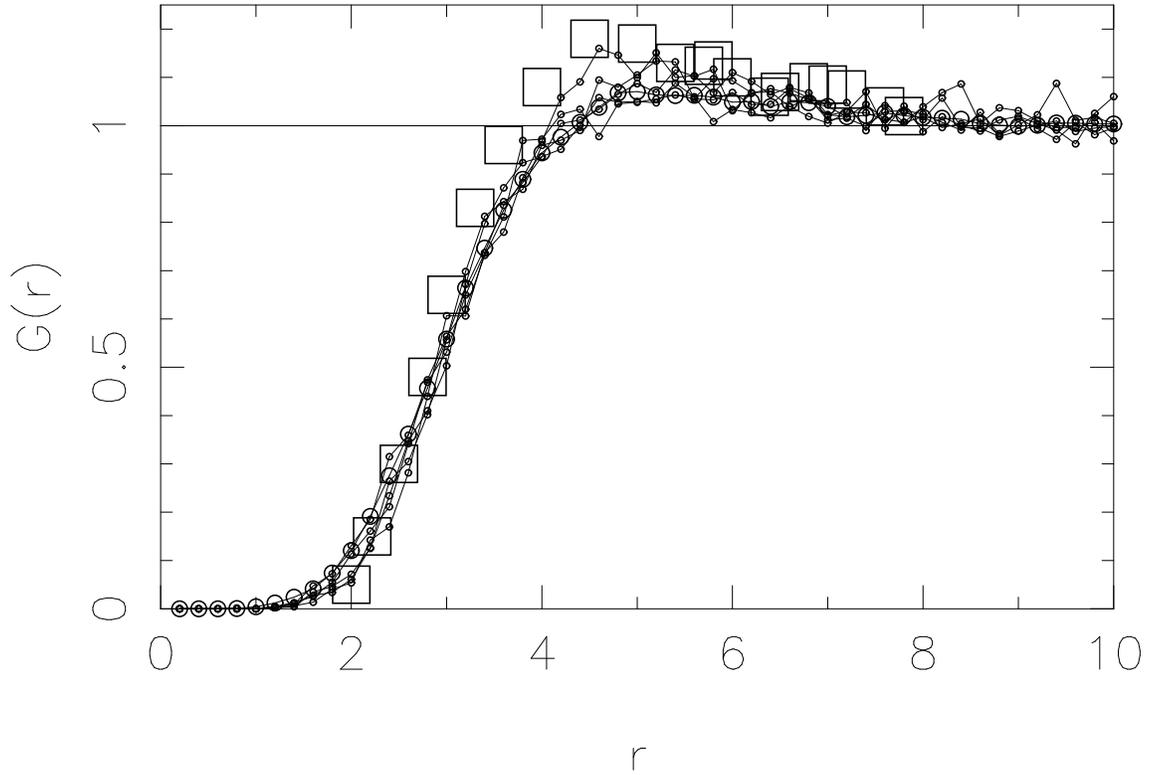}  }           
    \vspace{5mm}   
\caption{
Correlation functions of droplets' positions, $G(r)$, as measured at different
moments of time in the scaling state, corresponding to $N=3000, N=2000, N=1500,
N=1300, N=1000$ droplets present ({\it small circles}). The weighted average of
these runs (the weight of a configuration is proportional to the
corresponding number of droplets) is also shown ({\it large circles}). These
results were obtained using model B. Note that all the lines are
close to each other, indicating that system has indeed arrived at the scaling
state. The experimental results of Krichevsky and Stavans, shown for comparison
({\it squares}), are also close to ours.  
}
\end{figure}

\begin{figure}
\epsfxsize=140mm
        \centerline{ \epsffile{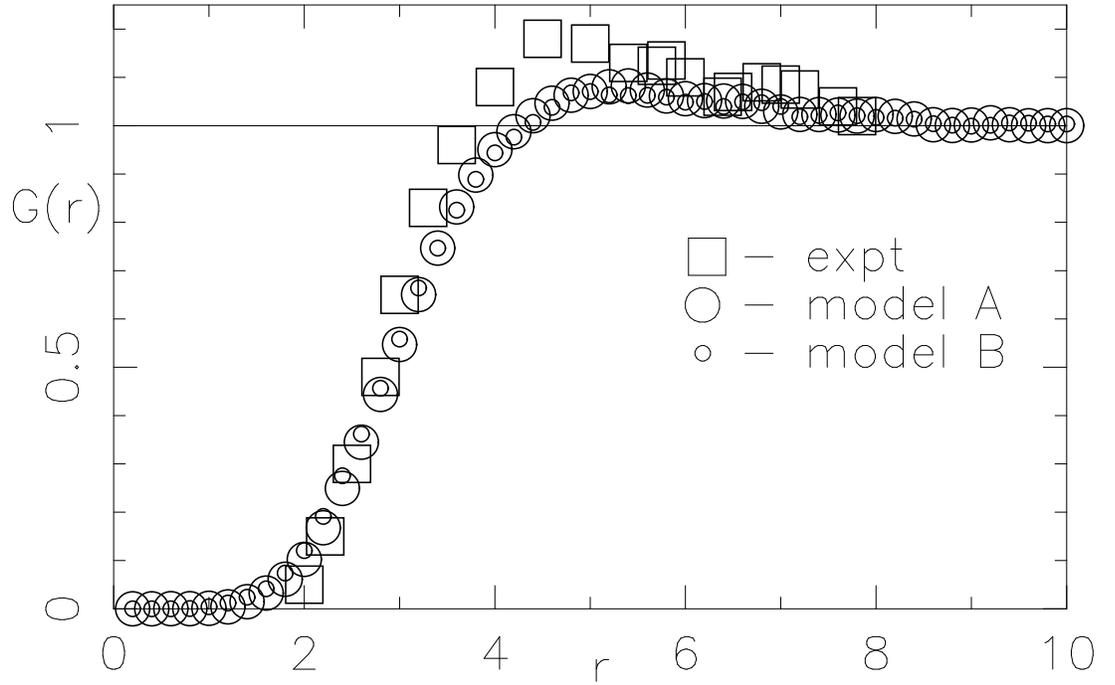}  }           
    \vspace{5mm}   
\caption{
Weighted time average (the weight of a configuration is proportional to its
number of droplets) of the correlation functions of droplets' positions,
$G(r)$,  (averaged over 8 runs) for models A and B in the scaling state ({\it
large circles} and {\it small circles} respectively). The experimental results
of Krichevsky and Stavans are also shown ({\it squares}). This shows that the
two models give indistinguishable results that are in good agreement with
experiment as well.
}
\end{figure}

\begin{figure}
\epsfxsize=140mm
        \centerline{ \epsffile{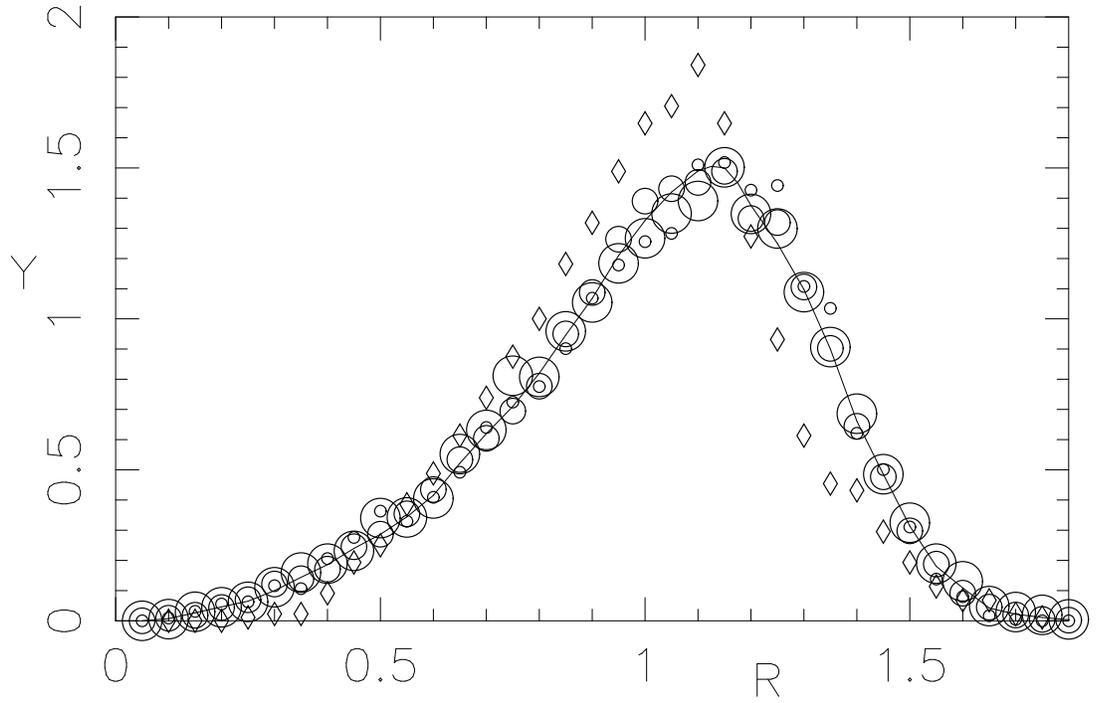}  }           
    \vspace{5mm}   
\caption{
$Y(R)$, the droplets' rescaled size distribution in the scaling state at 
$N=3000$ ({\it largest circles}), $N=2000$ ({\it intermediate circles} and
$N=1000$ ({\it small circles)}. Each plot presents the average over 8 runs. 
The line indicates the overall average. The experimental results of Krichevsky 
and Stavans are also shown ({\it diamonds}).  
}
\end{figure}

\begin{figure}
\epsfxsize=140mm
        \centerline{ \epsffile{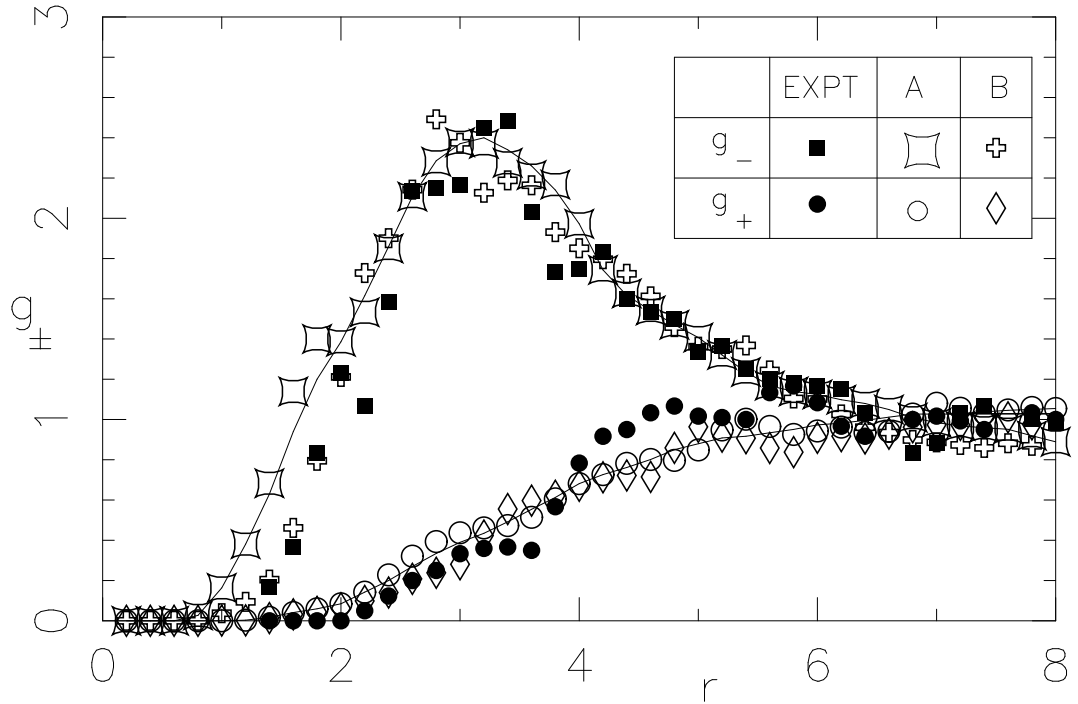}  }           
    \vspace{5mm}   
\caption{
The charge correlation functions (see precise definition in the text) for the
same ($g_+(r)$) and the opposite ($g_-(r)$) charges, as obtained using models A
and B, compared with the experimental data of Krichevsky and Stavans ({\it full
circles} and {\it full squares} for the same and the opposite charges
respectively). The lines are the guides for eye. Our data present averages over
8 runs. Note that the fluctuations in this plot are larger than for the
position correlation function (see Fig. 2). We believe that the difference
between the results of the two models is due to the fluctuations. 
Interestingly, model B seems to be closer to the experimental points.
}
\end{figure}



\begin{references}
\bibitem{NoteO}
When fluctuations in $R_j$ are neglected, this
equals $1/ \log (\langle R \rangle /R_0)$.
\bibitem{NoteI}
The role of the parameter $R_0$ in our theory is analogous to the parameter
$\Delta$ in Beenakker's model\cite{Been}; while being irrelevant for the exact 
problem, it was tuned in order to restore conservation of material, broken in 
the reduced description.
\end{references}
\end{document}